\documentclass[11pt,a4paper]{article}
\pdfoutput=1
\input xy
\xyoption{all}
\usepackage{amsmath}
\usepackage{amsthm}
\usepackage{amsfonts}
\usepackage{graphicx}
\usepackage{tikz}
\usepackage{pgfplots}

\usepackage{pict2e}
\usepackage{multirow}
\usepackage{colortbl}
\usepackage{color}

\usepackage{tabularx}
\newcolumntype{C}{>{\centering}X}

\usepackage{arydshln}
\definecolor{Gray}{gray}{0.9}
\definecolor{LightCyan}{rgb}{0.88,1,1}

\def\tp{\otimes}
\def\ds{\oplus}

\def\ket[#1]{\left| #1 \right>}
\def\bra[#1]{\left< #1 \right|}
\def\Bop[#1][#2][#3][#4][#5]{B\left(\left.\begin{array}{cc} #1&#2\\#3&#4\end{array}\right|#5\right)}

\def\R{\mathbb{R}}
\def\C{\mathbb{C}}
\def\L{\mathcal{L}}
\def\H{\mathcal{H}}

\theoremstyle{definition}

\theoremstyle{remark}

\title{The $D(D_{3})$-anyon chain: integrable boundary conditions\\ and excitation spectra}
\author{Peter E. Finch and Holger Frahm\\
  \\
  Institut f\"ur Theoretische Physik, \\
  Leibniz Universit\"at Hannover, \\
  Appelstra\ss e 2, 30167 Hannover, Germany}
\date{}

\begin{document}
% create title page
\maketitle

\begin{abstract}
  Chains of interacting non-Abelian anyons with local interactions invariant
  under the action of the Drinfeld double of the dihedral group $D_3$ are
  constructed. Formulated as a spin chain the Hamiltonians are generated from
  commuting transfer matrices of an integrable vertex model for periodic and
  braided as well as open boundaries. A different anyonic model with the same
  local Hamiltonian is obtained within the fusion path formulation.  This
  model is shown to be related to an integrable fusion interaction round the
  face model. Bulk and surface properties of the anyon chain are computed from
  the Bethe equations for the spin chain.  The low energy effective theories
  and operator content of the models (in both the spin chain and fusion path
  formulation) are identified from analytical and numerical studies of the
  finite size spectra.  For all boundary conditions considered the continuum
  theory is found to be a product of two conformal field theories.  Depending
  on the coupling constants the factors can be a $Z_4$ parafermion or a
  $\mathcal{M}_{(5,6)}$ minimal model.
\end{abstract} 

\setcounter{tocdepth}{2}
\tableofcontents

\section{Introduction}
In recent years there has been a surge of attention directed towards the
understanding of many-particle systems exhibiting topological order, i.e.
phases which cannot be characterised by a local order parameter.  Possible
realisations of such topological quantum liquids in condensed matter physics
are the fractional quantum Hall (FQH) states \cite{Laughlin1983,MooRea1991}
and certain two-dimensional frustrated quantum magnets
\cite{BaFiGi2002,Kitaev2006,MoeSon2001}. The excitations in these systems
display anyonic statistics and an understanding of their collective behaviour is
essential for the classification of topological phase transitions.
Particularly interesting are non-Abelian anyons where the interchange of two
particles is described by non-trivial representations of the braid group
complemented by fusion rules for the decomposition of product states.  The
fact that these non-Abelian anyons are protected by their topological charge
has led to proposals for the use of such systems in universal quantum
computation \cite{Kita03,NSSF08}.

Some insight into the peculiar properties of many interacting anyons can be
obtained in the context of simple model systems: such models can be obtained
by associating anyonic degrees of freedom with each site of a lattice and
defining interactions compatible with their braiding and fusion rules
\cite{FTLTKWF2007}.  The phase diagram of the resulting lattice models can be
studied based on the numerical computation of finite size spectra.  This
approach is particularly powerful for anyonic chains, i.e.\ one-dimensional
lattices, where the numerical data can be compared against predictions from
conformal field theory (CFT).  Another approach, also in one dimension, makes
use of the fact that the lattice model may become integrable for particular
choices of coupling constants.  For the solution of such models various
analytical methods, e.g.\ in the framework of the Quantum Inverse Scattering
Method (QISM), have been established which allow the study of the spectrum of
their low-energy excitations, their thermodynamical properties including the
long-distance asymptotics of correlation functions and even form factors
\cite{KKMST2011,KBIBook1993}.

So far much of the work on such lattice models has been focussed on systems of
the non-Abelian Ising or Fibonacci anyons related to the quasiparticles in
certain FQH states and their generalisations appearing in $su(2)_{k}$
Chern-Simons theories
\cite{AGLTT2011,FTLTKWF2007,GATLTW2009,LPTT2011,TAFHLT2008,TTWL2008}.  These
anyons have relatively simple fusion rules which allows for tractable
computation of systems with nearest and next-nearest neighbour interactions.
Integrable points within the one-dimensional versions of these models 
have been identified as restricted solid-on-solid (RSOS) or interaction round 
the face (IRF) models constructed from representations of Temperley-Lieb 
algebras \cite{FTLTKWF2007,KakArd2012,IkJS09}. An alternative
method to define an anyonic theory is via the Drinfeld doubles of a finite
group algebra, $D(G)$, and its representations \cite{WilBai1998}.  The
quasi-particles in these systems are irreducible representations (irreps) of
$D(G)$ labelled by their flux, i.e.\ an element of $h\in G$, and their
topological charge determined by the transformation properties under the
residual global symmetry commuting with the flux $h$.

Being a quasi-triangular Hopf algebra the quantum double allows for a direct
construction of integrable quantum chains with nearest neighbour interactions
described by a local Hamiltonian which is invariant under the corresponding
symmetry \cite{DIL2006,Finch2011}: within the QISM one obtains quantum spin
chains on a Hilbert space being a tensor product of the finite-dimensional
local spaces corresponding to a spin $S$, a qudit or a more general $n$-state
quantum system.  On the other hand, it is already known that for any given
model whose local Hamiltonian has the symmetry of a quasi-triangular Hopf
algebra associated with an anyonic theory, it is possible to construct quantum
chains using the fusion path formalism \cite{Finch2013}.  Here the basis
vectors are composed of sequences of anyons and we shall refer to this as a
fusion path chain.
The local Hamiltonian is formally identical in the spin and the fusion path
formalism.  Therefore, one should expect the bulk properties of the spin and
the fusion path model to be the same.  The finite size spectrum of low energy
excitations, however, is known to depend on boundary conditions
\cite{AlBB88,Card86b} and therefore should differ between the two
realisations.

In this paper we study this problem for a specific one-dimensional anyon chain
with nearest-neighbour interactions.  The underlying symmetry of the
Hamiltonian is that of the Drinfeld double of a dihedral group, specifically
$D(D_{3})$.  In the following section we define this algebra and recall its
irreps and the corresponding fusion rules.  Then, using the spin basis,
integrable models are constructed subject to periodic, braided and open
boundary conditions, all of which being based on the usual QISM transfer
matrix \cite{FFL2011}. 
While for the fusion path basis we construct a fusion IRF transfer matrix whose
series expansion contains the global one-dimensional Hamiltonian.
In Section~3 we compute the bulk and surface properties of the model from the
Bethe ansatz formulation of the spectral properties for the spin chain.  
The conformal field theory and operator content for the periodic spin chain
version of the $D(D_{3})$ model has been identified previously
\cite{FinFra2012}.  In Section~4 we expand this work providing more details on
the analysis as well as extending the study of the finite-size spectrum to the
spin chain with braided and open boundary conditions.  In addition we present 
results for the fusion path chain in support of the expectation that
the low energy excitations of the $D(D_3)$-anyon chain are described by the
same CFT for all types of boundary conditions studied here, namely products of
$Z_4$ parermion and $\mathcal{M}_{(5,6)}$ minimal models.

%%%%%%%%%%%%%%%%%%%%%%%%%%%%%%%%%%%%%%%%%%%%%%%%%%%%%%%%%%%%%%%%%%%%%%%%%%%%%%%%
%%%%%%%%%%%%%%%%%%%%%%%%%%%%%%%%%%%%%%%%%%%%%%%%%%%%%%%%%%%%%%%%%%%%%%%%%%%%%%%%
%%%%%%%%%%%%%%%%%%%%%%%%%%%%%%%%%%%%%%%%%%%%%%%%%%%%%%%%%%%%%%%%%%%%%%%%%%%%%%%%

\section{The Model and its Symmetries}
\subsection{The $D(D_{3})$ algebra}
The model we consider in this article has the underlying symmetry of the
Drinfeld (or quantum) double of a finite group algebra. The finite group we
utilise is the dihedral group of order six, $D_{3}$, and is isomorphic to the
group of permutations on three elements, $S_{3}$. This group is based upon the
symmetries of an equilateral triangle and has the presentation,
\begin{equation*}
	D_{3}  = \{\sigma, \tau | \sigma^{3} = \tau^{2} = \sigma \tau
        \sigma \tau = e \}\,, 
\end{equation*}
where $e$ is the identity element of the group, $\sigma$ is a rotation and
$\tau$ is a flip. The Drinfeld double of this group is defined as the vector
space, 
\begin{equation*}
	D(D_{3}) = \C \{gh^{*} | g,h \in D_{3}\},
\end{equation*}
where $*$ denotes an element from the dual space of $\C D_{3}$.
This space forms a quasi-triangular Hopf algebra when equipped with the
multiplication and coproduct, 
$$ 
  g_{1}h_{1}^{*} g_{2}h_{2}^{*} = \delta_{(h_{1}g_{2})}^{(g_{2}h_{2})} \,\,
  (g_{1}g_{2}) h_{2}^{*} \hspace{0.5cm} \mbox{and} \hspace{0.5cm} \Delta(gh^{*})
  = \sum_{k\in D_{3}} g (k^{-1}h)^{*} \tp g k^{*}.
$$
The remaining structure is uniquely determined by these relations
\cite{ChaPreBook1994,MajidBook1995}. This algebra has an associated universal
$R$-matrix i.e. an algebraic solution to the Yang--Baxter equation. \\

\noindent
\underline{\textbf{Representations}} \\
\noindent
The representation theory of the Drinfeld doubles of finite group algebras are
well known \cite{DPR1990,Gould1993}. The irreducible representations of
$D(D_{3})$ are classified by the conjugacy classes of $D_{3}$.
For a given conjugacy class a representative element is chosen and the
representations of the centraliser subgroup of this element are determined.
An action on conjugacy class is defined and then combined in prescribed manner
with an irrep of the centraliser. This yields an irrep of $D(D_{3})$ labelled
by both the representative element and the irrep of the centraliser.
The irreps associated with the conjugacy class $\{e\}$ are:
$$ \begin{array}{rclcrclcrcl}
  \pi^{(e,\pm)}(\sigma) & = & 1 && \pi^{(e,\pm)}(\tau) & = & \pm 1 &&
  \pi^{(e,\pm)}(g^{*}) & = & \delta_{g}^{e} \\ 
  \pi^{(e,1)}(\sigma) & = &
  \left(\begin{array}{cc}\omega&0\\0&\omega^{-1}\end{array}\right) &&
  \pi^{(e,1)}(\tau) & = &
  \left(\begin{array}{cc}0&1\\1&0\end{array}\right) &&
  \pi^{(e,1)}(g^{*}) & = &
  \left(\begin{array}{cc}\delta_{g}^{e}&0\\0&\delta_{g}^{e}\end{array} \right),
  \\ 
\end{array} 
$$
where $\omega=e^{\frac{2i\pi}{3}}$. The irreps associated with the conjugacy
class $\{\sigma,\sigma^{2}\}$ are:
$$ \begin{array}{rclcrclcrcl}
  \pi^{(\sigma,k)}(\sigma) & = &
  \left(\begin{array}{cc}\omega^{k}&0\\0&\omega^{-k}\end{array}\right) &&
  \pi^{(\sigma,k)}(\tau) & = &
  \left(\begin{array}{cc}0&1\\1&0\end{array}\right) && \pi^{(\sigma,k)}(g^{*})
  & = &
  \left(\begin{array}{cc}\delta_{g}^{\sigma}&0\\0&\delta_{g}^{\sigma^{2}} \end{array}\right),
  \\ 
\end{array}
$$
where $k \in \{0,1,2\}$.  The irreps associated with the conjugacy class
$\{\tau, \sigma\tau, \sigma^{2}\tau\}$ are:
$$ \begin{array}{rclcrclcrcl}
  \pi^{(\tau,\pm)}(\sigma) & = &
  \left(\begin{array}{ccc}0&0&1\\1&0&0\\0&1&0\end{array}\right) &&
  \pi^{(\tau,\pm)}(\tau) & = &
  \pm\left(\begin{array}{ccc}1&0&0\\0&0&1\\0&1&0\end{array}\right) &&
  \pi^{(\tau,\pm)}(g^{*}) & = &
  \left(\begin{array}{ccc}
      \delta_{g}^{\tau}&0&0\\0&\delta_{g}^{\sigma^{2}\tau}&0\\0&0&\delta_{g}^{\sigma\tau} \end{array}\right). \\
      \end{array} 
$$
The anyonic theory corresponding with $D(D_{3})$ associates an irrep with an
anyon \cite{WilBai1998}. For convenience it is simpler to denote each irreps
by a single letter, $\mathfrak{a},...,\mathfrak{h}$. We equate
$$
 \mathfrak{a}=(e,+),\, \mathfrak{b}=(e,-),\, \mathfrak{c}=(e,1),\,
 \mathfrak{d}=(\sigma,0),\, \mathfrak{e}=(\sigma,1),\,
 \mathfrak{f}=(\sigma,2),\, \mathfrak{g}=(\tau,+),\, \mathfrak{h}=(\tau,-) 
$$
Properties of the anyons are inherited from their associated irreps, e.g. the
dimension of an anyon equals the dimension of its corresponding irrep.\\

\noindent
\underline{\textbf{Fusion Rules}} \\
\noindent
Required for an anyonic theory are the fusion rules of particles. These rules
are defined by the tensor product decompositions of the associated irreps
\cite{Gould1993}:
\begin{eqnarray*}
  \pi^{\alpha} \tp \pi^{\beta}  =  \bigoplus_{\gamma} N_{\alpha\beta}^{\gamma} \pi^{\gamma}, 
  & \hspace{0.5cm} \mbox{where} \hspace{0.5cm} & 
  N_{\alpha\beta}^{\gamma} = \frac{1}{6} \sum_{g,h,k} \mbox{tr}[\pi^{\alpha}(h^{*}g^{-1})] \mbox{tr}[\pi^{\beta}(g (k^{-1}h)^{*})] \mbox{tr}[\pi^{\gamma}(g k^{*})]
\end{eqnarray*}
and tr is the trace. Applying the above formula yields the following fusion rules presented in terms of the associated labels:
\begin{center} \begin{tabular}{|c|ccc|ccc|cc|} \hline
$\tp$ & $\mathfrak{a}$ & $\mathfrak{b}$ & $\mathfrak{c}$ & $\mathfrak{d}$ & $\mathfrak{e}$ & $\mathfrak{f}$ & $\mathfrak{g}$ & $\mathfrak{h}$ \\ \hline
$\mathfrak{a}$ & $\mathfrak{a}$ & $\mathfrak{b}$ & $\mathfrak{c}$ & $\mathfrak{d}$ & $\mathfrak{e}$ & $\mathfrak{f}$ & $\mathfrak{g}$ & $\mathfrak{h}$ \\
$\mathfrak{b}$ & $\mathfrak{b}$ & $\mathfrak{a}$ & $\mathfrak{c}$ & $\mathfrak{d}$ & $\mathfrak{e}$ & $\mathfrak{f}$ & $\mathfrak{h}$ & $\mathfrak{g}$ \\
$\mathfrak{c}$ & $\mathfrak{c}$ & $\mathfrak{c}$ & $\mathfrak{a}\ds\mathfrak{b}\ds\mathfrak{c}$ & $\mathfrak{e}\ds\mathfrak{f}$ & $\mathfrak{d}\ds\mathfrak{f}$ & $\mathfrak{d}\ds\mathfrak{e}$ & $\mathfrak{g}\ds\mathfrak{h}$ & $\mathfrak{g}\ds\mathfrak{h}$ \\ \hline
$\mathfrak{d}$ & $\mathfrak{d}$ & $\mathfrak{d}$ & $\mathfrak{e}\ds\mathfrak{f}$ & $\mathfrak{a}\ds\mathfrak{b}\ds\mathfrak{d}$ & $\mathfrak{c}\ds\mathfrak{f}$ & $\mathfrak{c}\ds\mathfrak{e}$ & $\mathfrak{g}\ds\mathfrak{h}$ & $\mathfrak{g}\ds\mathfrak{h}$ \\
$\mathfrak{e}$ & $\mathfrak{e}$ & $\mathfrak{e}$ & $\mathfrak{d}\ds\mathfrak{f}$ & $\mathfrak{c}\ds\mathfrak{f}$ & $\mathfrak{a}\ds\mathfrak{b}\ds\mathfrak{e}$ & $\mathfrak{c}\ds\mathfrak{d}$ & $\mathfrak{g}\ds\mathfrak{h}$ & $\mathfrak{g}\ds\mathfrak{h}$ \\
$\mathfrak{f}$ & $\mathfrak{f}$ & $\mathfrak{f}$ & $\mathfrak{d}\ds\mathfrak{e}$ & $\mathfrak{c}\ds\mathfrak{e}$ & $\mathfrak{c}\ds\mathfrak{d}$ & $\mathfrak{a}\ds\mathfrak{b}\ds\mathfrak{f}$ & $\mathfrak{g}\ds\mathfrak{h}$ & $\mathfrak{g}\ds\mathfrak{h}$ \\ \hline
$\mathfrak{g}$ & $\mathfrak{g}$ & $\mathfrak{h}$ & $\mathfrak{g}\ds\mathfrak{h}$ & $\mathfrak{g}\ds\mathfrak{h}$ & $\mathfrak{g}\ds\mathfrak{h}$ & $\mathfrak{g}\ds\mathfrak{h}$ & $\mathfrak{a}\ds\mathfrak{c}\ds\mathfrak{d}\ds\mathfrak{e}\ds\mathfrak{f}$ & $\mathfrak{b}\ds\mathfrak{c}\ds\mathfrak{d}\ds\mathfrak{e}\ds\mathfrak{f}$ \\
$\mathfrak{h}$ & $\mathfrak{h}$ & $\mathfrak{g}$ & $\mathfrak{g}\ds\mathfrak{h}$ & $\mathfrak{g}\ds\mathfrak{h}$ & $\mathfrak{g}\ds\mathfrak{h}$ & $\mathfrak{g}\ds\mathfrak{h}$ & $\mathfrak{b}\ds\mathfrak{c}\ds\mathfrak{d}\ds\mathfrak{e}\ds\mathfrak{f}$ & $\mathfrak{a}\ds\mathfrak{c}\ds\mathfrak{d}\ds\mathfrak{e}\ds\mathfrak{f}$ \\ \hline
\end{tabular} \\ \end{center}

\subsection{Local Spin Hamiltonians}
The $D(D_3)$ model is constructed by taking a special case of the three state
Fateev--Zamolodchikov model \cite{FatZam1982b}. This limit yields the
$R$-matrix, which can also be constructed from the $\pi^{\mathfrak{g}} \tp
\pi^{\mathfrak{g}}$ representation of $D(D_{3})$ \cite{Finch2011},
\begin{equation}
 \label{eqnRmat}
 R(z_{1},z_{2}) = N(z_{1},z_{2})\sum_{a,b,i,j=0}^{2} \left[w^{(i-j)(a-b)}
   \overline{W}(z_{1}|a) \overline{W}(z_{2}^{-1}|b) \right] e_{i+a+b,i} \tp
 e_{j+a+b,j}\,,  
\end{equation}
where $e_{i,j}$ represents a $3 \times 3$ matrix (whose indices are considered
modulo three) with a one in the $i$th row and $j$th column and zeros
elsewhere,
$$ 
  \overline{W}(z|l) = \left[\frac{z - 1}{wz - w^{2}}\right]^{1-\delta_{l}^{0}}
  \hspace{0.7cm} \mbox{and} \hspace{0.7cm} N(z_{1},z_{2}) = -\frac{1}{3}(w z_{1}
  - w^{2}) (w - w^{2}z_{2}).
$$ 
The $R$-matrix satisfies a Yang-Baxter equation in \textit{both} the first and the second spectral parameter
\begin{eqnarray} 
\label{eqnYBE}
	R_{12}(x_{1},x_{2})R_{23}(x_{1}y_{1},x_{2}y_{2})R_{12}(y_{1},y_{2}) &
        = &
        R_{23}(y_{1},y_{2})R_{12}(x_{1}y_{1},x_{2}y_{2})R_{23}(x_{1},x_{2}), 
\end{eqnarray}
and has the symmetry of $D(D_{3})$, implying that the operator can be
expressed in terms of projection operators. The projection operators from 
$\pi_{\mathfrak{g}} \tp \pi_{\mathfrak{g}}$ to the irreps in its decomposition
are,
\begin{equation}
\label{proj3x3}
	p^{(a)} = \frac{\dim{(a)}}{6} \sum_{g,h}
        \mbox{tr}[\pi^{a}(h^{*}g^{-1})] (\pi^{\mathfrak{g}} \tp
        \pi^{\mathfrak{g}})\Delta(gh^{*}). 
\end{equation}
In terms of these projection operators the $R$-matrix is written as,
\begin{eqnarray} \label{RmatProj}
	R(z_{1},z_{2}) & = & f_{\mathfrak{a}}(z_{1},z_{2})p^{(\mathfrak{a})} +
        f_{\mathfrak{c}}(z_{1},z_{2}) p^{(\mathfrak{c})} +
        f_{\mathfrak{d}}(z_{1},z_{2})p^{(\mathfrak{d})} +
        f_{\mathfrak{e}}(z_{1},z_{2})p^{(\mathfrak{e})} +
        f_{\mathfrak{f}}(z_{1},z_{2})p^{(\mathfrak{f})},
\end{eqnarray}
where
$$ 
  f_{a}(z_{1},z_{2})p^{(a)} = R(z_{1},z_{2})p^{(a)}, \hspace{2cm} a \in
  \{\mathfrak{a},\mathfrak{c},\mathfrak{d},\mathfrak{e},\mathfrak{f} \}.
$$ 
This $R$-matrix allows us to construct integrable models subject to various
boundary conditions \cite{FFL2011}.  In the spin chain formulation each
lattice site carries a representation $\pi^{\mathfrak{g}}$ of $D(D_3)$.  As a
consequence of the dependence of the $R$-matrix on two spectral parameters
there exist two local Hamiltonians describing the interaction between
neighbouring spins in the Hilbert space from $\pi^{\mathfrak{g}} \tp
\pi^{\mathfrak{g}}$ representation.  The local Hamiltonians are obtained in
the usual manner by taking derivatives of the $R$-matrix with respect to the
spectral parameters:
\begin{equation*}
	h^{(k)} = i \left.\frac{d}{dz_{k}} R(z_{1},z_{2})\right|_{z_1=1,z_2=1}
        - \beta_{k} I \tp I\,, 
\end{equation*}
where $k\in\{1,2\}$ and $\beta_{k}\in\C$ is chosen such that the trace of the
local Hamiltonians is zero.  In terms of the projectors (\ref{proj3x3})
the local Hamiltonians are given by \cite{FinFra2012}
\begin{equation}
\label{eqnLocalHam}
\begin{aligned}
  h^{(1)} & = \frac{2\sqrt{3}}{3}p^{(\mathfrak{a})} -
  \frac{\sqrt{3}}{3}p^{(\mathfrak{c})} - \frac{\sqrt{3}}{3}p^{(\mathfrak{d})}
  - \frac{\sqrt{3}}{3}p^{(\mathfrak{e})} +
  \frac{2\sqrt{3}}{3}p^{(\mathfrak{f})}\,,\\
  h^{(2)} & = \frac{2\sqrt{3}}{3}p^{(\mathfrak{a})} -
  \frac{\sqrt{3}}{3}p^{(\mathfrak{c})} - \frac{\sqrt{3}}{3}p^{(\mathfrak{d})}
  + \frac{2\sqrt{3}}{3}p^{(\mathfrak{e})} -
  \frac{\sqrt{3}}{3}p^{(\mathfrak{f})}\,.
\end{aligned}
\end{equation}
  
It follows that the local Hamiltonians commute with each other and with the
action of the algebra:
$$ 
\left[ h^{(1)}, h^{(2)} \right] = 0 \hspace{0.5cm} \mbox{and} \hspace{0.5cm}
\left[ (\pi_{\mathfrak{g}} \tp \pi_{\mathfrak{g}})\Delta(a), h^{(k)} \right] =
0,
$$
for all $a \in D(D_{3})$.  Therefore they have the underlying symmetry of
$D(D_{3})$ as the $R$-matrix did. From explicit calculation of $h^{(1)}$ and
$h^{(2)}$ we find,
$$ 
h^{(1)} = \Pi h^{(2)}\Pi = \left[h^{(2)}\right]^{*} 
$$
where $\Pi$ is usual two-site permutation operator and $*$ is, and herein
reserved for, complex conjugation.  Both local Hamiltonians
(\ref{eqnLocalHam}) are self-adjoint.

\subsection{Global Hamiltonian}
In the following we shall consider a variety of models with interactions
described by the local operators (\ref{eqnLocalHam}) but subject to different
boundary conditions.  As a consequence of the existence of two distinct local
Hamiltonians the global Hamiltonian is comprised of two terms weighted by a
free coupling parameter $\theta\in[0,2\pi]$ as
\begin{equation} 
  \label{eqnCompleteHam}
  \mathcal{H}_{\theta} = \cos(\theta)\,\mathcal{H}^{(1)} +
     \sin(\theta)\,\mathcal{H}^{(2)}\,. 
\end{equation}
For all boundary conditions considered below these models are integrable
thanks to the existence of a commuting transfer matrix.  Furthermore the two
components of the global Hamiltonian will commute,
\begin{equation} \label{eqnCommutingComponents}
	\left[\mathcal{H}^{(1)}, \mathcal{H}^{(2)}\right] = 0.
\end{equation}
This commutativity will be particularly useful in the investigation of the
models as it allows us to sutdy the spectra of $\mathcal{H}^{(1)}$ and
$\mathcal{H}^{(2)}$ separately.  Typically the spectra of $\mathcal{H}^{(1)}$
and $\mathcal{H}^{(2)}$ will be identical or of a related form.

\subsubsection{Periodic Spin Chain}
We begin by considering the $D(D_3)$ model as a spin chain with periodic
boundary conditions: its global Hamiltonians are defined by
\begin{equation*}
  \mathcal{H}^{(k)}  = h^{(k)}_{\mathcal{L} 0} + \sum_{j=1}^{\mathcal{L}-1} h^{(k)}_{j(j+1)},
\end{equation*}
for $k \in \{1,2\}$.  Note that the periodic closure by the term
$h^{(k)}_{\mathcal{L} 0}$ in the global Hamiltonian breaks the $D(D_{3})$
invariance of the model.
Both of these Hamiltonians appear in the series expansion of the commuting
transfer matrix,
\begin{equation*}
  t(z_{1},z_{2}) = \mbox{tr}_{0} \left[\Pi_{0\mathcal{L}}R_{0\mathcal{L}}(z_{1},z_{2}^{*})
    \cdots \Pi_{01}R_{01}(z_{1},z_{2}^{*}) \right]\,.  
\end{equation*}
By construction this transfer matrix is a polynomial of degree $\mathcal{L}$
in the variables $z_1$ and $z_2^*$.  It has been observed that this transfer
matrix factorises and that its eigenvalues are always of the form
\cite{FFL2011}
\begin{equation}
 \label{eqnTranferEig}
 \Lambda(z_{1},z_{2}) = c
 \prod_{\ell=1}^{\mathcal{L}}(z_{1}-z_{1,\ell})\prod_{\ell=1}^{\mathcal{L}}
 (z_{2}-z_{2,\ell})^{*}\,. 
\end{equation}
Therefore the eigenvalues can be conveniently described in terms of their
zeroes $z_{k,\ell} \equiv i\omega \mathrm{e}^{x_{k,\ell}}$, for $k=1,2$ and
$\ell=1,\ldots,\mathcal{L}$.
Furthermore, starting from the $D(D_{3})$ fusion rules functional relations
satisfied by the transfer matrices (or equivalently their eigenvalues) can be
derived \cite{Finch2011}:
\begin{equation}
  \label{eqnFunRelPeriodic}
  \begin{aligned}
  \lambda_{1}(z_{1}) \Lambda(z_{1},z_{2}) & = (\omega z_{1}+1)^{\mathcal{L}}
  \Lambda(\omega z_{1},z_{2}) + (z_{1}-1)^{\mathcal{L}} \Lambda(\omega^{-1}
  z_{1},z_{2}), \\ 
  \lambda_{2}(z_{2}) \left[\Lambda(z_{1},z_{2})\right]^{*} & = (\omega
  z_{2}+1)^{\mathcal{L}} \left[\Lambda(z_{1},\omega z_{2})\right]^{*} +
  (z_{2}-1)^{\mathcal{L}} \left[\Lambda(z_{1},\omega^{-1} z_{2}))\right]^{*},
\end{aligned}
\end{equation}
where $\lambda_{k}(z)$ are analytic functions.  This implies that the two
sets of parameters $\{x_{k,\ell}\}_{\ell=1}^{\mathcal{L}}$ must independently
satisfy the Bethe equations \cite{FFL2011}
\begin{equation}
  \label{eqnBAperiodic}
  (-1)^{\mathcal{L}+1} \left(\frac{1+(i/\omega) \mathrm{e}^{x_{k,j}}}{ 1-i\omega\,
      \mathrm{e}^{x_{k,j}}}\right)^{\mathcal{L}}  
  =  \prod_{l=1}^{\mathcal{L}} \frac{\mathrm{e}^{x_{k,l}}
    -(1/\omega)\mathrm{e}^{x_{k,j}}}{\mathrm{e}^{x_{k,l}}
    -\omega\,\mathrm{e}^{x_{k,j}}}\,,\quad j=1,\ldots,\mathcal{L}\,. 
\end{equation}
It is important to note that while there are exactly $\mathcal{L}$ Bethe roots
in each set $\{x_{k,\ell}\}_{\ell=1}^{\mathcal{L}}$, they are allowed to be at
$\pm \infty$, but at most one at each.
The energy eigenvalue of $\H^{(k)}$ corresponding to the set of Bethe roots
$\{x_{k,\ell}\}$ is given by
\begin{equation}
 \label{eqnEnCom} 
 E^{(k)} \equiv E(\{x_{k,\ell}\})  =  i \left[ \sum_{\ell=1}^{\mathcal{L}}
   \frac{1}{1-i\omega e^{x_{k,\ell}}} - \frac{1}{6}\left(3 + i
     \sqrt{3}\right) \mathcal{L} 
 \right]\,.
\end{equation}
Here we have used the property that sets of Bethe roots are invariant under
complex conjugations, $\{x_{k,\ell}\}_{\ell=1}^{\mathcal{L}} \equiv
\{x_{k,\ell}^{*}\}_{\ell=1}^{\mathcal{L}} \pmod{2i\pi}$.  Since the local
Hamiltonians are Hermitian by construction, the energies (\ref{eqnEnCom}) must
be real. This reality of the energy imposes an additional physicality
constraint on solutions to the Bethe equation (care must be taken to deal with
roots at $\pm\infty$ appropriately).  We note that all the root configurations
considered below in the discussion of the spectrum of the system do satisfy
this condition.

Let us remark that the Bethe equations (\ref{eqnBAperiodic}) and the
corresponding energies (\ref{eqnEnCom}) of the $D(D_3)$ spin chain of even
length $\mathcal{L}$ coincide with those of the three-state Potts spin chain
with $\mathcal{L}/2$ sites \cite{AlDM92}. We shall use this equivalence below
to identify some of the thermodynamical properties of the $D(D_3)$ chain.

The energy eigenvalues of the \emph{complete} Hamiltonian are characterised by
two solutions to the Bethe equations (\ref{eqnBAperiodic}).  As a consequence
of Equations (\ref{eqnCompleteHam}), (\ref{eqnCommutingComponents}) and along
with (\ref{eqnEnCom}) they are given by
\begin{equation}
\label{eqnEtot}
	E  = \cos(\theta)\,E^{(1)} + \sin(\theta)\,E^{(2)} =
        \cos(\theta)\,E(\{x_{1,\ell}\}) + \sin(\theta)\,E(\{x_{2,\ell}\}), 
\end{equation}
provided these energies (or the corresponding root configurations) pair.
Specifically, levels are said to pair if the two corresponding sets of Bethe
roots form an eigenvalue of the transfer matrix, see Equation
(\ref{eqnTranferEig}). As the two sets of Bethe roots need not correspond to a
unique eigenvalue of the transfer matrix, e.g. there may be two eigenvalues
that differ by a constant factor or an eigenvalue might be degenerate, we
refer to the total number of eigenvalues, including degeneracies, as the
\textit{pairing multiplicity}.

The total momentum of the corresponding state can also be given in terms of
the two sets of Bethe roots: at $z_{1}=1=z_{2}$ the transfer matrix becomes a
shift operator by one site.  Therefore the momentum operator is $P=-i\ln
t(1,1)$.  By construction the eigenvalues of this operator are real
($2\pi/\mathcal{L}$ times an integer for periodic boundary conditions
considered here).
Unlike for the energy (\ref{eqnEnCom}) it is not possible to identify
\emph{partial} momentum contribution from one of the participating Bethe
configurations uniquely \cite{FFL2011}.
Using the invariance of the sets of Bethe roots under complex conjugation we
use
\begin{equation}
\label{eqnMoCom}
  P^{(k)} \equiv P(\{x_{k,\ell}\}) = \mbox{Re}\left[\frac{1}{i}
    \sum_{\ell=1}^{\mathcal{L}} \ln(1-i\omega e^{x_{k,\ell}})\right]
  =  \frac{1}{2i}\,\sum_{\ell=1}^{\mathcal{L}} \ln\left(
    \frac{1-i\omega\mathrm{e}^{x_\ell}}{1-(1/i\omega)\mathrm{e}^{x_\ell}}\right)
  \,.
\end{equation}
as definition of the partial momenta $P^{(k)}$.  For later use we note that
the second expression is \emph{half} of the momentum of the three-state Potts
spin chain \cite{AlDM92}.  Consistency with Eq.~(\ref{eqnTranferEig}) implies
that the \emph{complete} momentum is related to the partial ones as
\begin{equation}
  \label{eqnMotot}
  P  = P^{(1)} - P^{(2)} + \mathrm{const.}\,
\end{equation}
Again, roots $x_{k,\ell}=\pm\infty$ have to be taken into account to ensure finite
(partial) momentum.  The total momentum is given by the difference of partial
momenta reflecting the fact $\mathcal{H}^{(2)}$ is the spatial inversion of
$\mathcal{H}^{(1)}$.  The remaining constant represents a macroscopic
effect, details of which have been discussed in earlier works \cite{FFL2011}.

\subsubsection{Braided Chain}
One closed chain proposed as an alternative to the periodic chain is the
braided chain \cite{Foerster1996,GPPR1994,KarZap1994}.  In this model,
translational invariance is replaced by invariance under a global braiding
operator.  As a consequence the underlying symmetry of the model will not be
broken, i.e.\ it has the full global $D(D_{3})$ symmetry. The global
Hamiltonians for these boundary conditions are defined by,
$$
  \mathcal{H}^{(k)} = Bh^{(k)}_{(\mathcal{L}-1)\mathcal{L}}B^{-1} +
  \sum_{j=1}^{\mathcal{L}-1} h^{(k)}_{j(j+1)}, \hspace{1cm} k \in \{1,2\}, 
$$ 
where
$$
  B = b_{12}b_{23}...b_{(\mathcal{L}-1)\mathcal{L}}, \hspace{1cm} \mbox{and}
  \hspace{1cm} b = \lim_{z\rightarrow\infty} \left[ \frac{1}{z^{2}} R(z,z)
  \right].  
$$
There also exist different possible definitions for the braiding operator 
$b_{i}$ relating to other limits of $R(z_{1},z_{2})$ \cite{Finch2011}.
The integrability of the braided model is ensured by the existence of a
transfer matrix $t(z_{1},z_{2})$, which can be found in \cite{Finch2011}.
Eigenstates of this model are again characterised by the Bethe Equations
(\ref{eqnBAperiodic}).  As in the periodic case there must be $\mathcal{L}$
Bethe roots, this time with one Bethe root allowed at $+\infty$ but none
allowed at $-\infty$.

As mentioned above the Hamiltonian of this model is invariant under the action
of the global braiding operator.  Specifically, we find that $B$ can be
realized by the transfer matrix of the braided model as $B=t(1,1)$.
Furthermore, the braiding operator is idempotent: from the analysis of small
systems we find that
\begin{eqnarray} \label{eqnOrderB}
  B^{n} & = & I, \quad \mbox{when} \,\, \left\{\begin{array}{ccc} n = 3\mathcal{L},
     & & \mathcal{L}\,\, \mbox{even},\\ n = 2\mathcal{L}, & & \mathcal{L}\,\,
     \mbox{odd}. \end{array} \right.
\end{eqnarray}
This allows us to define an analog of the momentum operator as the generator
of braiding operations by $P_b = -i\ln B$.  The eigenvalues of $P_b$ are
restricted to integer multiples of either $\frac{\pi}{\mathcal{L}}$ or
$\frac{2\pi}{3\mathcal{L}}$ depending on the parity of $\mathcal{L}$.

\subsubsection{Open Boundary Conditions}
\label{SecOpenIntro}
We also consider spin chains with open boundary conditions.  In this case
integrable models derive from representations of Sklyanin's reflection algebra
\cite{Sklyanin1988}.  $c$-number representations of this algebra define
possible boundary terms.  For the present model these $K$-matrices are found
to be the same as have been determined for the $D(D_{3})$ one-parameter
$R$-matrix \cite{DFIL2009}.  The global Hamiltonians are,
\begin{equation*}
  \mathcal{H}^{(k)}  = \chi_{k}^{-} B_{1}^{(k)-} + \chi_{k}^{+}
  B_{\mathcal{L}}^{(k)+} + \sum_{j=1}^{\mathcal{L}-1} h^{(k)}_{j(j+1)}\,. 
\end{equation*}
Here, the boundary operators $B^{(k)-}$ ($B^{(k)+}$) act on the first (last)
site of the chain, respectively.  There exist three possible (independent)
options for each of these operators, namely
\begin{eqnarray} \label{eqnBfields}
  B^{(1)-}, (B^{(1)+})^{*}, (B^{(2)-})^{*}, B^{(2)+} & \in & 
  \left\{\left.\left( 
  \begin{array}{ccc}	
    0 & \omega^{2}b & \omega^{2}b^{2} \\ 
    \omega b^{2} & 0 & b \\ 
    \omega b & b^{2} & 0 \end{array} \right) 
   \right| b=1,\omega,\omega^{2} \right\}.
\end{eqnarray}
The real boundary amplitudes where $\chi_{k}^{\pm}$, $k\in \{1,2\}$, have to
satisfy $\chi_{1}^{+} \chi_{2}^{+} = \chi_{1}^{-} \chi_{2}^{-} = 0$.  Like the
periodic and braided models integrability is derived from the existence of a
transfer matrix (see \cite{Finch2011,FFL2011} for the open $D(D_{3})$ transfer
matrix) and the eigenstates of the Hamiltonian are classified by sets of Bethe
roots.  The Bethe equations for the Hamiltonian $\mathcal{H}^{(k)}$ are
independent of the choice of the boundary operators, $B^{(k)\pm}$,
\begin{equation}
\label{eqnBAopen}
\begin{aligned}
  \prod_{l=1}^{d_{k}}
  \left(\frac{e^{x_{k,l}}-\omega^{2}e^{x_{k,j}}}{e^{x_{k,l}}-\omega
      e^{x_{k,j}}}\right)  
  =&  (-1)^{\mathcal{L}+1} \left( \frac{1+\omega
      e^{2x_{k,j}}}{1+\omega^{2}e^{2x_{k,j}}} \right) \left(
    \frac{1-\omega^{2}e^{2x_{k,j}}}{1-\omega e^{2x_{k,j}}} \right)
  \left( \frac{1 + i\omega^{2}e^{x_{k,j}}}{1-i\omega e^{x_{k,j}}}
  \right)^{2\mathcal{L}}  \\ 
  & \times \Phi(x_{k,j},\chi_{k}^{-}) \Phi(x_{k,j},\chi_{k}^{+})
\end{aligned}
\end{equation}
with Bethe roots always appearing in pairs of $\pm x$ and where
\begin{equation*}
  \Phi(x,\chi) =  \left\{ \begin{array}{ccl}
      1 & \mathrm{for~}\chi = 0, \\ 
      \left(\frac{\chi\sqrt{3}(1-\omega e^{2x})+
          i\omega^{2}\left(1+\chi\sqrt{3}\right)e^{x}}{
          \chi\sqrt{3}(1-\omega^{2}e^{2x})  
          - i\omega \left(1+\chi\sqrt{3}\right) e^{x}}\right)
      & \mathrm{for~}\chi \neq 0. \end{array}\right. 
\end{equation*}
In addition to the explicit dependence of the Bethe equations on the boundary
amplitudes $\chi_{k}^{\pm}$ the latter determine the number $d_k$ of Bethe
roots $x_{k,j}$: 
for the open chain with free ends, i.e.\ $\chi_{k}^{\pm} = 0$, there are
$d_{k} = 2\mathcal{L}$ Bethe roots with at most one pair of roots at
$\pm\infty$ \cite{FFL2011}.
A non-zero boundary term at one end of the chain, i.e.\ choosing one of either
$\chi_{k}^{+}$ or $\chi_{k}^{-}$ non-zero, changes the number of roots to
$d_{k} = 2\mathcal{L}+2$, while for boundary terms at both ends there are
$d_{k}=2\mathcal{L} + 4$ Bethe roots.  We refer to the extra pairs of roots
appearing as compared to the free-ends case as boundary roots.\footnote{It is
  important to note that for $\chi_{k}^{\pm}$ small the presences of the
  boundary Bethe roots will not have a significant effect on the configuration
  of the bulk Bethe roots.}
These boundary roots are finite for non-zero boundary amplitudes
$\chi_{k}^{\pm}$ but approach $\pm\infty$ in the limit of free-ends.

The energy of the open boundary Hamiltonian $\H^{(k)}$ corresponding to a
solution of Eqs.~(\ref{eqnBAopen}) is given by
\begin{equation*}
  E^{(k)}(\{x_{k,\ell}\}) =
  \frac{i}{2}\left\{\sum_{l=1}^{d_{k}}\left[\frac{1}{1-i\omega e^{x_{k,l}}}
    \right] -(1-i\frac{\sqrt{3}}{3})\mathcal{L} - \phi(\chi_{k}^{+}) -
    \phi(\chi_{k}^{-})\right\}, 
\end{equation*}
where
\begin{equation*}
  \phi(\chi) = \left\{ \begin{array}{ccl}
     0 & \mathrm{for~}\chi = 0\,, \\ 
  1+i\chi & \mathrm{for~}\chi \neq 0\,. \end{array}\right.
\end{equation*}
Again we note that the energy eigenvalues are real as the global Hamiltonian
is Hermitian.  Additionally, it is important to note that the presence of
boundary interactions breaks the $D(D_{3})$ invariance of the model.  Only for
free ends the open model has this invariance.

\subsection{Fusion Path Analogues}
As the local Hamiltonians have the symmetry of $D(D_{3})$ it is possible to
create fusion path analogues \cite{Finch2013}. Depending on the boundary
conditions imposed the global Hamiltonians may or may not be equivalent to
their spin formalism counterparts discussed above.  The construction of the
analogous fusion path chains uses the Pasquier's method of representation
theory reliant face-vertex correspondence \cite{Pasquier1988}.  This allows
the fusion path analogues to be considered as the Hamiltonian limits of RSOS
models and proves their integrability.  The connection between fusion IRF
models and many other physical systems has already been established
\cite{Gepner1992}.

We first define the fusion path basis. Basis vectors of the fusion path space
are of the form,
$$ \left| a_{0}a_{1}...a_{\mathcal{L}}\right>, $$
where $a_{i} \in \{\mathfrak{a},\mathfrak{b},\mathfrak{c},
\mathfrak{d},\mathfrak{e},\mathfrak{f},\mathfrak{g},\mathfrak{h}\}$ and
neighbouring labels satisfy the condition,
\begin{eqnarray}
  a_{i}a_{i+1} 
  & \in & \{ab \,|\, V_{b} \subset V_{a} \tp V_{\mathfrak{g}} \} \label{EqnAllowedNeighbour} \\
  & = & \{\mathfrak{a}\mathfrak{g}, \mathfrak{b}\mathfrak{h},
  \mathfrak{c}\mathfrak{g},\mathfrak{c}\mathfrak{h},
  \mathfrak{d}\mathfrak{g},\mathfrak{d}\mathfrak{h},
  \mathfrak{e}\mathfrak{g},\mathfrak{e}\mathfrak{h},
  \mathfrak{f}\mathfrak{g},\mathfrak{f}\mathfrak{h}, \nonumber \\  
  && \phantom{\{} \mathfrak{g}\mathfrak{a}, \mathfrak{g}\mathfrak{c},
  \mathfrak{g}\mathfrak{d}, \mathfrak{g}\mathfrak{e},
  \mathfrak{g}\mathfrak{f}, \mathfrak{h}\mathfrak{b},
  \mathfrak{h}\mathfrak{c}, \mathfrak{h}\mathfrak{d},
  \mathfrak{h}\mathfrak{e}, \mathfrak{h}\mathfrak{f}\}. \nonumber 
\end{eqnarray}
Thus $a_{i+1}$ must appear in the fusion of $a_{i}$ and
$\mathfrak{g}$. Diagrammatically a basis vector corresponds to the figure
below, where the joining of two lines indicates fusion which occurs from left
to right and top to bottom, \vspace{0.5cm}
\begin{center}
\begin{tikzpicture}[scale=1.0]
	%Anyon positions
	\put (30,0){$\mathfrak{g}$}	\put (60,0){$\mathfrak{g}$}	\put (90,0){$\mathfrak{g}$}	\put (150,0){$\mathfrak{g}$}	\put (180,0){$\mathfrak{g}$}
	\put (0,-20){$a_{0}$}	\put (40,-27){$a_{1}$}	\put (70,-27){$a_{2}$}	\put (100,-27){$a_{3}$}	\put (160,-27){$a_{\mathcal{L}-1}$}	\put (205,-20){$a_{\mathcal{L}}$}
	%Bulk line
	\draw (0.5,-0.6) -- (4.0,-0.6);	\draw (4.2,-0.6) -- (4.4,-0.6);	\draw (4.6,-0.6) -- (4.8,-0.6);	\draw (5.0,-0.6) -- (7.0,-0.6);
	%Vertical lines
	\draw (1.15,-0.1) -- (1.15,-0.6);	\draw (2.20,-0.1) -- (2.20,-0.6);	\draw (3.25,-0.1) -- (3.25,-0.6);	\draw (5.35,-0.1) -- (5.35,-0.6);	\draw (6.40,-0.1) -- (6.40,-0.6);
\end{tikzpicture}\vspace{0.1cm}
\end{center}
To construct local operators on this space we utilise $F$-moves (generalised
6-j symbols), which allow the temporary re-ording of fusion, \vspace{0.3cm}
\begin{eqnarray*}
\begin{tikzpicture}[scale=1.0]
	%Anyon positions
	\put (26,5){$b$}	\put (56,5){$c$}
	\put (0,-20){$a$}	\put (40,-27){$d$}	\put (83,-20){$e$}
	%Bulk line
	\draw (0.35,-0.6) -- (2.85,-0.6);
	%Vertical lines
	\draw (1.00,0.1) -- (1.00,-0.6);	\draw (2.05,0.1) -- (2.05,-0.6);
\end{tikzpicture} \hspace{0.3cm} 
& = & 
\sum_{d'} (F^{abc}_{e})^{d}_{d'} \hspace{0.5cm}
\begin{tikzpicture}[scale=1.0]
	%Anyon positions
	\put (26,5){$b$}	\put (56,5){$c$}
	\put (8,-20){$a$}	\put (47,-15){$d'$}	\put (76,-20){$e$}
	%Bulk line
	\draw (0.60,-0.6) -- (2.60,-0.6);
	%Vertical lines
	\draw (1.00,0.1) -- (1.525,-0.25);	\draw (2.05,0.1) -- (1.525,-0.25); 	\draw (1.525,-0.25) -- (1.525,-0.6);
\end{tikzpicture}
\end{eqnarray*}
In terms of these $F$-moves we can define two-site projection operators which
act non-trivially on a single link $i$ of the fusion path lattice\footnote{It
  is important for the reader to note that in this fusion path formalism 
  the labels in the basis vectors do not correspond to individual sites but
  rather bonds. The individual sites are still the $\mathfrak{g}$-anyons but
  now can not be solely acted on as this would break local $D(D_{3})$
  invariance.},
\begin{equation*}
  \tilde{p}^{(b)}_{i-1,i,i+1}
  = \sum_{a_{i-1},a_{i},a_{i}',a_{i+1}}
  \left[\left(F^{a_{i-1}\mathfrak{g}\mathfrak{g}}_{a_{i+1}}\right)^{a_{i}'}_{b}\right]^{*}
  \left(F^{a_{i-1}\mathfrak{g}\mathfrak{g}}_{a_{i+1}}\right)^{a_{i}}_{b}
  \left|..a_{i-1}a_{i}'a_{i+1}..\right>\left<..a_{i-1}a_{i}a_{i+1}..\right|. 
\end{equation*}
The unitary $F$-moves associated with $D(D_{3})$ can be calculated explicitly
from the representation theory of $D(D_{3})$ \cite{RomersPrivate}.
Below we shall consider local interactions defined analogously to the nearest
neighbour spin chain ones, Eq.~(\ref{eqnLocalHam}):
\begin{equation*}
\begin{aligned}
  \tilde{h}^{(1)} & = \frac{2\sqrt{3}}{3}\tilde{p}^{(\mathfrak{a})} -
  \frac{\sqrt{3}}{3}\tilde{p}^{(\mathfrak{c})} -
  \frac{\sqrt{3}}{3}\tilde{p}^{(\mathfrak{d})} -
  \frac{\sqrt{3}}{3}\tilde{p}^{(\mathfrak{e})} +
  \frac{2\sqrt{3}}{3}\tilde{p}^{(\mathfrak{f})} \\ 
  \tilde{h}^{(2)} & = \frac{2\sqrt{3}}{3}\tilde{p}^{(\mathfrak{a})} -
  \frac{\sqrt{3}}{3}\tilde{p}^{(\mathfrak{c})} -
  \frac{\sqrt{3}}{3}\tilde{p}^{(\mathfrak{d})} +
  \frac{2\sqrt{3}}{3}\tilde{p}^{(\mathfrak{e})} -
  \frac{\sqrt{3}}{3}\tilde{p}^{(\mathfrak{f})} 
\end{aligned}
\end{equation*}
Note that these Hamiltonians act on three consecutive labels of the fusion
path basis but only can change the middle label.

As a consequence of the equivalence of the local interactions between the spin
and fusion path formalisms the global models in the two formalisms may differ
only by boundary conditions.
The open model with free ends and braided model both have $D(D_{3})$
invariance which means that the fusion path and spin versions of these chains
are equivalent. This implies that the energy spectra are identical and the
denegeracies appearing in each formalism are related via a mapping. For
convenience we use the degeneracy of the spin chain formalism for these
choices of boundary conditions.
For periodic boundary conditions, however, neither the spin chain nor fusion
path model has the complete $D(D_{3})$ invariance and thus the two models,
while sharing bulk properties, are distinct \cite{Finch2013}.  

To construct a periodic model in the fusion path basis we need to consider the
space spanned by the basis vectors satisfying $a_{0}=a_{\mathcal{L}}$. As a
consequence of the fusion rules this periodic closure is possible only for
lattices of even length $\mathcal{L}$.  Furthermore, they lead to the
decomposition of the Hilbert space
\begin{equation}
  \mbox{Hilbert Space}  = \C\left\{\left| a_{1}...a_{\mathcal{L}}\right> |
    a_{1}=\mathfrak{g} \, \mbox{or} \, a_{1}=\mathfrak{h} \right\} \oplus
  \C\left\{\left| a_{1}...a_{\mathcal{L}}\right> | a_{2}=\mathfrak{g} \,
    \mbox{or} \, a_{2}=\mathfrak{h} \right\}, 
\label{EqnFPHSdecomp} 
\end{equation}
where each of these subspaces has dimension $3^{\mathcal{L}}+1$. The global
Hamiltonians are,
$$ 
  \tilde{\mathcal{H}}^{(k)} = \tilde{h}^{(k)}_{(\mathcal{L}-1)\mathcal{L} 1} +
  \tilde{h}^{(k)}_{\mathcal{L} 1 2} + \sum_{j=2}^{\mathcal{L}-1}
  \tilde{h}^{(k)}_{(j-1)j(j+1)}, \hspace{1cm} k \in \{1,2\}. 
$$

The integrability of this model can be established based on the existence of
an $R$-matrix connected to an RSOS model whose heights correspond to the
labels of the irreps of $D(D_{3})$.  As the $R$-matrix of Eq.~(\ref{RmatProj})
is expressible in terms of the $D(D_{3})$ projection operators it follows that
there exists an equivalent operator in the fusion path basis
\cite{Pasquier1988},
\begin{equation}
 \label{RmatProjFusion}
 \begin{aligned}
   \tilde{R}(z_{1},z_{2})
   & =  f_{\mathfrak{a}}(z_{1},z_{2})\tilde{p}^{(\mathfrak{a})} +
   f_{\mathfrak{c}}(z_{1},z_{2}) \tilde{p}^{(\mathfrak{c})} +
   f_{\mathfrak{d}}(z_{1},z_{2}) \tilde{p}^{(\mathfrak{d})} +
   f_{\mathfrak{e}}(z_{1},z_{2}) \tilde{p}^{(\mathfrak{e})} +
   f_{\mathfrak{f}}(z_{1},z_{2})
   \tilde{p}^{(\mathfrak{f})}\\ 
   & = \sum_{a_{1},a_{2},a_{2}',a_{3}}
   \Bop[a_{1}][a_{2}'][a_{2}][a_{3}][z_{1},z_{2}]
   \ket[a_{1}a_{2}'a_{3}]\bra[a_{1}a_{2}a_{3}],
 \end{aligned}
\end{equation}
satisfying a face Yang-Baxter equation
\begin{equation*} 
\label{eqnYBEFusion}
  \tilde{R}_{123}(x_{1},x_{2}) \tilde{R}_{234}(x_{1}y_{1},x_{2}y_{2})
  \tilde{R}_{123}(y_{1},y_{2}) =
  \tilde{R}_{234}(y_{1},y_{2}) \tilde{R}_{123}(x_{1}y_{1},x_{2}y_{2})
  \tilde{R}_{234}(x_{1},x_{2}),
\end{equation*}
The weights appearing in $\tilde{R}(z_{1},z_{2})$ are used to construct the
commuting transfer matrix \cite{GRS1996},
\begin{equation*}
  \bra[a_{1}'...a_{\mathcal{L}}']\tilde{t}(z_{1},z_{2})\ket[a_{1}...a_{\mathcal{L}}] 
  =  \prod_{j=1}^{\mathcal{L}}
  \Bop[a_{j}'][a_{j+1}'][a_{j}][a_{j+1}][z_{1},z_{2}]. 
\end{equation*}
Once again this transfer matrix generates a family of commuting operators,
including the global Hamiltonians $\tilde{\mathcal{H}}^{(1)}$ and
$\tilde{\mathcal{H}}^{(2)}$, implying integrability. Again, the dependence of the
transfer matrix on two spectral parameters guarantees commutativity of the
two components
$$ 
\left[\tilde{\mathcal{H}}^{(1)}, \tilde{\mathcal{H}}^{(2)} \right] = 0. 
$$
Through the analysis of small finite size systems we are able to make two
important observations.  Firstly, we find that the eigenvalues
$\tilde{\Lambda}(z_{1},z_{2})$ of the fusion path transfer matrix factorize
into two polynomials of degree $\mathcal{L}$ in the same manner as those in the
spin chain case (\ref{eqnTranferEig}).  Secondly, we find that the eigenvalues
satisfy functional relations similar to the Eqs.~(\ref{eqnFunRelPeriodic}),
i.e.
\begin{equation} 
  \label{eqnFunRelFusPath}
  \begin{aligned}
  \lambda_{1}(z_{1}) \tilde{\Lambda}(z_{1},z_{2}) & =  (\omega
  z_{1}+1)^{\mathcal{L}} \tilde{\Lambda}(\omega z_{1},z_{2}) \pm
  (z_{1}-1)^{\mathcal{L}} \tilde{\Lambda}(\omega^{-1} z_{1},z_{2}) \\ 
  \lambda_{2}(z_{2}) \left[\tilde{\Lambda}(z_{1},z_{2})\right]^{*} & =
  (\omega z_{2}+1)^{\mathcal{L}} \left[\tilde{\Lambda}(z_{1},\omega
    z_{2})\right]^{*} \pm (z_{2}-1)^{\mathcal{L}}
  \left[\tilde{\Lambda}(z_{1},\omega^{-1} z_{2}))\right]^{*}\,.
\end{aligned}
\end{equation}
Again, $\lambda_{k}(z)$ are analytic functions.  The $\pm$ sign depends upon
the eigenvalue in question.  Preliminary calculations indicate that this
relation can be obtained using the fusion procedure for RSOS models
\cite{BazRes1989,KluPea1992}.  Like the periodic spin chain case these
functional relations lead to Bethe Equations which have to be satisfied by the
zeroes of the transfer matrix eigenvalues,
\begin{equation}
  \label{eqnBAFusPath}
  (-1)^{\mathcal{L}+1} \left(\frac{1+(i/\omega) \mathrm{e}^{x_{k,j}}}{ 1-i\omega\,
      \mathrm{e}^{x_{k,j}}}\right)^{\mathcal{L}}  
  =  \eta \prod_{l=1}^{\mathcal{L}} \frac{\mathrm{e}^{x_{k,l}}
    -(1/\omega)\mathrm{e}^{x_{k,j}}}{\mathrm{e}^{x_{k,l}}
    -\omega\,\mathrm{e}^{x_{k,j}}}\,,\quad j=1,\ldots,\mathcal{L}\,, 
\end{equation}
where $\eta=\pm1$ corresponds to the allowed signs in the functional relations
(\ref{eqnFunRelFusPath}).  This sign was not present in the Bethe equations
for the spin chain (\ref{eqnBAperiodic}) indicating the likely presence of
different excitations.  As before, every set of roots $\{x_{k,\ell}\}$ solving
the Bethe equations (\ref{eqnBAFusPath}) parametrizes an eigenvalue of the
fusion path model.  The energy eigenvalue of $\tilde{\mathcal{H}}^{(k)}$ is
again given by Equation (\ref{eqnEnCom}).

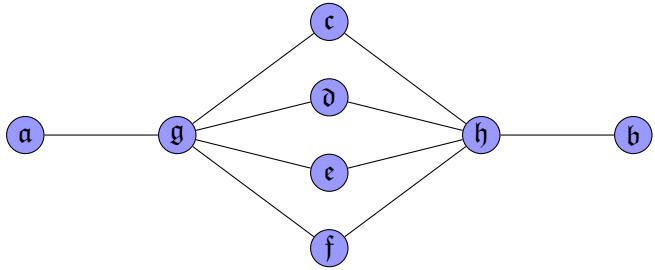
\begin{figure}[ht]
\caption{A graphical representation of allowed neighbouring labels in the
  fusion path chain. The vertices/nodes of the graphs are the labels of anyons
  which are connected via an edge if and only if the two anyon labels can
  appear next to each as given by Equation
  (\ref{EqnAllowedNeighbour}).\label{FigAllowedNeighbours}} 
\begin{center}
\begin{tikzpicture}[scale=1.0]
	\tikzstyle{every node}=[circle,draw,thin,fill=blue!40,minimum size=14pt,inner sep=0pt]
	\node (na) at (0,0) {$\mathfrak{a}$};
	\node (nb) at (8.0,0) {$\mathfrak{b}$};
	\node (nc) at (4.0,1.5) {$\mathfrak{c}$};
	\node (nd) at (4.0,0.5) {$\mathfrak{d}$};
	\node (ne) at (4.0,-0.5) {$\mathfrak{e}$};
	\node (nf) at (4.0,-1.5) {$\mathfrak{f}$};
	\node (ng) at (2.0,0) {$\mathfrak{g}$};
	\node (nh) at (6.0,0) {$\mathfrak{h}$};
	\foreach \from/\to in
        {ng/na,ng/nc,ng/nd,ng/ne,ng/nf,nh/nb,nh/nc,nh/nd,nh/ne,nh/nf} 
	\draw (\from) -- (\to);
\end{tikzpicture}
\end{center}
\end{figure}

In previous studies of anyonic fusion path models
\cite{FTLTKWF2007,KakArd2012} integrability was also observed by relating them
to transfer matrices associated with RSOS models.  However, in these instances
the fusion path $R$-matrices corresponded to representations of the
Temperley-Lieb algebra.
Every RSOS model can be naturally associated with a graph where nodes
represent the labels of anyons in the theory which are connected if they can
appear next to each other in the fusion path basis as given by
Eq.~(\ref{EqnAllowedNeighbour}), see \cite{Pasquier1987a}.  For the $D(D_{3})$
model considered here we obtain the graph given in Figure
\ref{FigAllowedNeighbours}.  This graph is equivalent to McKay's
representation graph for the representation $\pi_{\mathfrak{g}}$ of $D(D_{3})$
\cite{McKay1980}.  We note that this graph shows that the $D(D_{3})$ fusion
path model does not correspond to any of the known RSOS models
associated with Dynkin diagrams
\cite{Pasquier1987a,Pasquier1987b,Roche1992,WarNie1993}.  It also does not
appear amongst the more general graphs associated with other RSOS models
\cite{FraZub1990}.

%%%%%%%%%%%%%%%%%%%%%%%%%%%%%%%%%%%%%%%%%%%%%%%%%%%%%%%%%%%%%%%%%%%%%%%%%%%%%%%%
%%%%%%%%%%%%%%%%%%%%%%%%%%%%%%%%%%%%%%%%%%%%%%%%%%%%%%%%%%%%%%%%%%%%%%%%%%%%%%%%
%%%%%%%%%%%%%%%%%%%%%%%%%%%%%%%%%%%%%%%%%%%%%%%%%%%%%%%%%%%%%%%%%%%%%%%%%%%%%%%%

\section{The Bethe Equations and Exact Results for Spin Chains}
As a consequence of Eq.~(\ref{eqnEtot}) the ground state energy of the model
is always obtained by the following combinations (NB: provided that these
states are allowed to pair) \cite{FFL2011}
\begin{equation} \label{GSEForm}
	E_{0} =
	\left\{ \begin{array}{lrcl} 
\cos(\theta)E_{l}^{(1)} + \sin(\theta)E_{l}^{(2)}, & 0 \leq & \theta & <
\frac{\pi}{2}\,, \\ 
\cos(\theta)E_{h}^{(1)} + \sin(\theta)E_{l}^{(2)}, & \frac{\pi}{2} \leq &
\theta & < \pi\,, \\ 
\cos(\theta)E_{h}^{(1)} + \sin(\theta)E_{h}^{(2)}, & \pi \leq & \theta & <
\frac{3\pi}{2}\,, \\ 
\cos(\theta)E_{l}^{(1)} + \sin(\theta)E_{h}^{(2)}, & \frac{3\pi}{2} \leq &
\theta & < 2\pi\,.
\end{array} \right.	
\end{equation}
Here $E_{l}^{(k)}$ is the lowest energy of the $\mathcal{H}^{(k)}$ and
$E_{h}^{(k)}$ is the highest.  
An immediate implication of this form of the ground state energy is that there
are level-crossings for $\theta$ being integer multiples of $\frac{\pi}{2}$
leading to first order quantum phase transitions.  
Here we will use a different consequence of (\ref{GSEForm}): the complete
spectrum of the model can be obtained from an analysis at these particular
points in combination with the implementation of the pairing rules
\cite{FFL2011}.  An additional simplification arises from the fact that the
spectra of $\mathcal{H}^{(1)}$ and $\mathcal{H}^{(2)}$ are identical for most
boundary conditions considered in this paper: this allows us to restrict the
analysis of the low-energy spectrum to those of the Hamiltonians
$\mathcal{H}_{\theta=0}=\mathcal{H}^{(1)}$ and $\mathcal{H}_{\theta=\pi}=
-\mathcal{H}^{(1)}$ whose ground state energies are $E_{l}^{(1)}=E_{l}^{(2)}$
and $-E_{h}^{(1)}=-E_{h}^{(2)}$, respectively.
Only in the case of open boundary conditions with $D(D_3)$-symmetry breaking
boundary fields the spectra of $\pm\mathcal{H}^{(1)}$ and
$\pm\mathcal{H}^{(2)}$ are independent.

\subsection{Energy Density in the Thermodynamical Limit} \label{SecEnergyDense}
The study of the excitation spectrum of a model requires knowledge of the bulk 
properties. As such we recall previously obtained results and present them here 
for completeness \cite{FFL2011}.
In the thermodynamic limit $\mathcal{L}\to\infty$ bulk properties of the
system are independent of the boundary conditions imposed. Therefore we can
compute the energy density from the Bethe equations (\ref{eqnBAperiodic}) for
the periodic spin chain.  To this end the solutions to the Bethe equations
need to be classified and the root configurations corresponding to the ground
state and low energy excitations have to the identified.  As mentioned above,
the Bethe equations for the periodic $D(D_3)$ spin chain arise also in the
context of the 3-state Potts model.  For the latter the classification of
solutions has been obtained by Albertini \emph{et al.} \cite{AlDM92,AlDM92a},
see also Ref.~\cite{FFL2011}.
In particular, numerical diagonalisation of the transfer matrix shows that the
lowest energy states of $\H_{\theta=\pi}$ consist of three different Bethe
root ($z_{l}\equiv z_{1,l}$) types: 
\begin{enumerate}
\item Positive Bethe roots ($+$-string), $z_{l}=i\omega e^{x_{l}^{+}}$, where
  $x_{l}^{+}\in \R$, 
\item Negative Bethe roots ($-$-string), $z_{l}=i\omega e^{x_{l}^{-}+i\pi}$,
  where $x_{l}^{-}\in \R$, 
\item 2-strings, where the Bethe roots come in pairs, $z_{l}=i\omega
  e^{x_{l}^{s}+\frac{2i\pi}{3}}, z_{l+1}=i\omega
  e^{x_{l}^{s}-\frac{2i\pi}{3}}$ with $x_{l}^{s}\in\R$, 
\end{enumerate}
as well as a limited number of Bethe roots at $\pm \infty$. Letting $N_{+}$,
$N_{-}$ and $N_{2}$ be the number of $+$-strings, $-$-strings and $2$-strings
respectively and setting $n_{\pm\infty} \in \{0,1\}$ to be the number of Bethe
roots at $\pm\infty$ then we have the constraint
$$ N_{+} + N_{-} + 2N_{s} = \mathcal{L} - n_{+\infty} - n_{-\infty}. $$
In the 3-states Potts model these root types were also identified, along with
a few other which we don't consider, with the additional constraint
$n_{+\infty}=n_{-\infty}$ \cite{AlDM92}.

Allowing combinations of these roots we then find that the for the periodic
Hamiltonian we can take the logarithm of the Bethe equations and define the
following set of counting functions:
\begin{eqnarray*}
  Z_{+}(x) & = & - \phi(x;\frac{7}{12}) +
  \frac{1}{\mathcal{L}}\sum_{l=1}^{N_{+}}\phi(x-x_{l}^{+};\frac{1}{3}) +
  \frac{1}{\mathcal{L}}\sum_{l=1}^{N_{-}}\phi(x-x_{l}^{-};\frac{5}{6}) +
  \frac{1}{\mathcal{L}}\sum_{l=1}^{N_{s}}\phi(x-x_{l}^{s};\frac{2}{3})
  \\ 
  Z_{-}(x) & = & - \phi(x;\frac{1}{12}) +
  \frac{1}{\mathcal{L}}\sum_{l=1}^{N_{+}}\phi(x-x_{l}^{+};\frac{5}{6}) +
  \frac{1}{\mathcal{L}}\sum_{l=1}^{N_{-}}\phi(x-x_{l}^{-};\frac{1}{3}) +
  \frac{1}{\mathcal{L}}\sum_{l=1}^{N_{s}}\phi(x-x_{l}^{s};\frac{1}{6})
  \\ 
  Z_{s}(x) & = & \left[\phi(x;\frac{11}{12}) + \phi(x;\frac{1}{4})\right] -
  \frac{1}{\mathcal{L}}\sum_{k=1}^{N_{+}} \phi(x-x_{l}^{+};\frac{2}{3}) -
  \frac{1}{\mathcal{L}}\sum_{l=1}^{N_{-}}\phi(x-x_{l}^{-};\frac{1}{6}) -
  \frac{1}{\mathcal{L}}\sum_{l=1}^{N_{s}}\phi(x-x_{l}^{s};\frac{1}{3}) 
\end{eqnarray*}
where
$$ 
 \phi(x;t) = -\frac{1}{\pi} \tan^{-1}
 \left(\frac{\tanh(\frac{x}{2})}{\tan(t\pi)}\right). 
$$
In the thermodynamical limit we find that the ground state for
$\H_{\theta=\pi}$ consists entirely of 2-strings \cite{AlDM92,FFL2011}. For
finite size systems this configuration is only realised when $\mathcal{L}$ is
even. The lowest energy Bethe root configuration for odd $\mathcal{L}$ is
given by $\frac{\mathcal{L}-1}{2}$ 2-strings and one $\pm$-string.

Similarly we find that the lowest energy states of $\H_{\theta=0}$ consist of
the same three Bethe root types and hence we have the same counting functions.
In the thermodynamical limit the Bethe root configuration of the ground state
consists of only negative and positive Bethe roots appearing in the ratio of
three $-$-strings to one $+$-string.  For finite size systems this
configuration is only realised when $\mathcal{L}$ is a
multiple of four.  The lowest energy Bethe root configuration for the other
chain lengths also consists of only negative and positive Bethe roots
appearing approximately in the ratio 3:1.

Based on these observations the root density formalism \cite{YaYa69} can be
applied to compute the corresponding energy densities: the density of
$2$-strings in the thermodynamic ground state of $\H_{\theta=\pi}$ and their
dressed energies $\H_{\theta=\pi}$ are determined by linear integral equations
\begin{equation}
\label{eqndressedPI}
\begin{aligned}
	\rho(x) & = \frac{1}{\pi}\left(\frac{1}{4\cosh(x) - 2\sqrt{3}} -
          \frac{1}{2\cosh(x)}\right) +
        \frac{\sqrt{3}}{2\pi}\int_{-\infty}^{\infty} \mathrm{d}y\,
            \frac{1}{2\cosh(x-y)+1} \rho(y)\,, \\ 
	\epsilon(x) & = \frac{1}{4\cosh(x) - 2\sqrt{3}} -
        \frac{1}{2\cosh(x)} + 
        \frac{\sqrt{3}}{2\pi}\int_{-\infty}^{\infty}\mathrm{d}y\,
        \frac{1}{2\cosh(x-y)+1} \epsilon(y)\,. 
\end{aligned}
\end{equation}
These equations (and the corresponding ones for $\mathcal{H}_{\theta=0}$) can be solved
by Fourier transformation giving the ground state energy densities
\cite{Hamer81,AlDM92,FFL2011}
\begin{equation}
  \frac{1}{\L}\, E_{\theta=\pi} = -\left[ \frac{1}{\pi} + \frac{2\sqrt{3}}{9}
  \right] 
\hspace{1cm} \mbox{and} \hspace{1cm} 
  \frac{1}{\L}\, E_{\theta=0} =
-\left[\frac{1}{2\pi}-\frac{2\sqrt{3}}{9}+\frac{3}{4}\right]\,. 
\end{equation}

\subsection{Fermi-velocity}
The low-energy excitations over these ground states have a linear dispersion
and their Fermi velocities have been computed within the root density
formalism in the context of the three-state Potts model \cite{AlDM92}.
As discussed above, it is possible to identify energy and momentum eigenvalues
of this model with those of the partial Hamiltonians $\mathcal{H}^{(k)}$
(\ref{eqnEnCom}) and corresponding momenta (\ref{eqnMoCom}) using the
equivalence of the corresponding Bethe equations.  

Noting that  the contribution of a single 2-string to the partial momentum can
be expressed in terms of their density in the thermodynamic limit
\begin{equation}
\label{eqndressedmom}
  p(x) = \pi \int^x \mathrm{d}y\, \rho(y)
\end{equation}
we can elimininate the rapidity $x$ from Eqs.~(\ref{eqndressedPI}) and
(\ref{eqndressedmom}) to obtain the dispersion relation $\epsilon(p)$ of
2-strings.  Therefore the Fermi velocity of low lying excitations of
$\mathcal{H}_{\theta=\pi}$ is found to be
\begin{equation}
 \label{eqnvFPI}
  v_F = \left.\frac{\partial \epsilon(p)}{\partial p}\right|_{p=p_F}
  = \left. \frac{1}{\pi} \frac{\epsilon'(x)}{\rho(x)} \right|_{x=-\infty}
  = 3 
\end{equation}
in agreement with the finite size analysis of the spectrum performed in
\cite{FinFra2012}. 

Similarly we can compute the Fermi velocity of gapless excitations for the
Hamiltonian $\mathcal{H}_{\theta=0}$.  Again the result is twice than what has been found for
the three-state Potts chain \cite{AlDM92}, i.e.\ $v_{F} = \frac{3}{2}$.

\subsection{Boundary Fields}
We can also determine the exact expressions for the surface energy, i.e. the
$\mathcal{L}^{0}$ contributions to the energy, for the open model with
interacting boundary fields. Firstly we find the ground energy for the
Hamiltonians $\mathcal{H}_{\theta=\pi}$ and $\mathcal{H}_{\theta=0}$, and then
extend this result to generic $\theta$ using Eq.~(\ref{GSEForm}).  Starting
with the Bethe equations (\ref{eqnBAopen}) we can apply the same method that
was used to calculate bulk energy density in Section \ref{SecEnergyDense}.
For the open spin chain with free ends the surface energy has been computed
previously \cite{FFL2011}.  Due to the symmetry of the Bethe equations the
general case can be studied in the context of the open chain with a single
boundary field present, e.g.\ $\chi_1^+\ne0$ and all other boundary amplitudes
vanishing, and compute the correction to case of the free ends.

For even $\mathcal{L}$ and $\chi_{1}^{+}$ not too large the Bethe root
configuration corresponding to the ground state of $\mathcal{H}_{\theta=\pi}$
consists of $\mathcal{L}$ 2-strings, distributed symmetrically around the
imaginary axis (just as in the free-ends case \cite{FFL2011}) and, in addition
two boundary Bethe roots.  The latter are found to be either $\pm$-strings
depending on the sign of $\chi_{1}^{+}$.
As the magnitude of $\chi_{1}^{+}$ is decreased these boundary Bethe roots tend
towards $\pm\infty$ to recover the Bethe configurations of the free-ends
model.\footnote{On a technical note, it was observed the magnitude of the
  boundary Bethe roots will also increase as $\mathcal{L}$ increases for fixed
  $\chi$. We assume that as $\mathcal{L}$ goes to $\infty$ the boundary Bethe
  roots tend to $\pm\infty$ for fixed boundary amplitudes $\chi_k^\pm$. This
  is important when considering the thermodynamical limit.}
We find that the correction to surface energy for the Hamiltonian
$\mathcal{H}_{\theta=\pi}$ with one interacting boundary field, compared to
the free ends case, is
\begin{equation*}
  g_{\theta=\pi}(\chi) 
  =  -\frac{2\chi^{2}\sqrt{3}}{1+2\chi\sqrt{3}} -
  \frac{18\chi^{3}\sqrt{3}}{\pi (1+2\chi\sqrt{3})
    \sqrt{1+2\chi\sqrt{3}-9\chi^{2}}}  
  \left\{ \begin{array}{ll} \mbox{arccosh}(-\frac{1}{2} -
      \frac{1}{2\chi\sqrt{3}}), & \chi < 0, \\  
      0, & \chi = 0, \\
      \mbox{arccosh}(\frac{1}{2} + \frac{1}{2\chi\sqrt{3}}), & \chi >
      0. \end{array} \right.
\end{equation*}
Similarly, we find the surface energy correction for $\mathcal{H}_{\theta=0}$,
\begin{equation*}
  g_{\theta=0}(\chi)
  = \frac{2\chi^{2}\sqrt{3}}{1+2\chi\sqrt{3}} - 
  \left\{ \begin{array}{ll}
      \frac{-9\chi\sqrt{-\chi}}{2(1+2\chi\sqrt{3})\sqrt{\sqrt{3}-3\chi}}  +
      \frac{9\chi^{3}\sqrt{3}
        \mbox{arccosh}(-\frac{1}{2}-\frac{1}{2\chi\sqrt{3}})}{\pi 
        (1+2\chi\sqrt{3}) \sqrt{1+2\chi\sqrt{3}-9\chi^{2}}}, & \chi < 0, \\ 
      0, & \chi = 0, \\
      \frac{9\chi\sqrt{\chi}}{2\sqrt{\sqrt{3}+9\chi}}  +
      \frac{9\chi^{3}\sqrt{3}
        \mbox{arccosh}(\frac{1}{2}+\frac{1}{2\chi\sqrt{3}})}{\pi 
        (1+2\chi\sqrt{3}) \sqrt{1+2\chi\sqrt{3}-9\chi^{2}}}, & \chi >
      0. \end{array} \right. 
\end{equation*}
Formally, the calculation of these corrections requires $-\frac{1}{3\sqrt{3}}
< \chi < \frac{1}{\sqrt{3}}$.  The restriction to this interval is due to the
change in the analytical behaviour of the Bethe equations which is reflected
by the presence of poles in the above expressions.

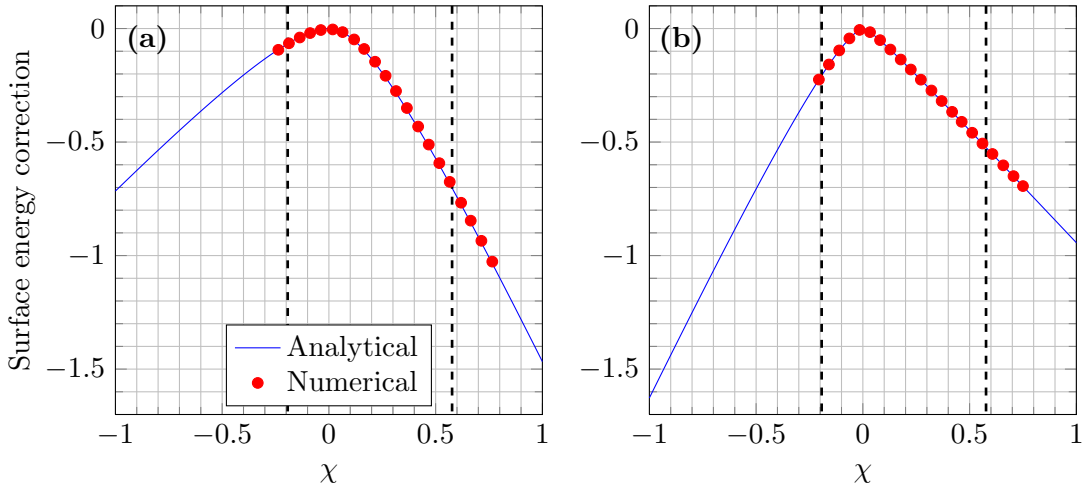
\begin{figure}[ht] 
  \caption{Numerical values of the correction to the surface energies for the
    Hamiltonians (a) $\mathcal{H}_{\theta=\pi}$ and (b)
    $\mathcal{H}_{\theta=0}$ as a function of the boundary amplitude $\chi$
    obtained solving the Bethe equations for a system of 100 sites compared to
    the analytical predictions $g_{\theta=\pi}(\chi)$ and
    $g_{\theta=0}(\chi)$, respectively. Dashed lines indicate the range,
    $-\frac{1}{3\sqrt{3}} < \chi < \frac{1}{\sqrt{3}}$, where the former
    expressions were formally derived. \label{figBoundaryCorrection}}
\begin{center}
\begin{tikzpicture}
	%\begin{axis}[width=7.2cm,height=7cm, grid=both, xmin=-1,xmax=1,ymin=-1.7,ymax=0.1, minor tick num=4, xlabel=$\chi$,ylabel=$g_{\theta=\pi}(\chi)$,title={Surface energy correction for $\mathcal{H}_{\theta=\pi}$}, legend style={at={(0.5,0.02)},anchor=south}]
	\begin{axis}[width=7.2cm,height=7cm, grid=both, xmin=-1,xmax=1,ymin=-1.7,ymax=0.1, minor tick num=4, xlabel=$\chi$,ylabel=Surface energy correction, legend style={at={(0.5,0.02)},anchor=south}]
		\addplot[smooth,blue] 
			coordinates {(-1,-0.71587) (-0.95,-0.67041) (-0.9,-0.62529) (-0.85,-0.58054) (-0.8,-0.5362) (-0.75,-0.49232) (-0.7,-0.44896)
			(-0.65,-0.4062) (-0.6,-0.36411) (-0.55,-0.32281) (-0.5,-0.28241) (-0.45,-0.24306) (-0.4,-0.20495) (-0.35,-0.16832) 
			(-0.3,-0.13347) (-0.25,-0.1008) (-0.2,-0.070836) (-0.15,-0.044327) (-0.1,-0.02234) (-0.05,-0.0065485) (0,0) (0.05,-0.0098687)
			(0.1,-0.03816) (0.15,-0.080736) (0.2,-0.134) (0.25,-0.19533) (0.3,-0.26285) (0.35,-0.33521) (0.4,-0.41142) (0.45,-0.49073)
			(0.5,-0.57258) (0.55,-0.65653) (0.6,-0.74223) (0.65,-0.82943) (0.7,-0.91789) (0.75,-1.0074) (0.8,-1.0979) (0.85,-1.1892) 
			(0.9,-1.2813) (0.95,-1.3739) (1,-1.4671)}; \addlegendentry{Analytical}		
		\addplot[only marks,red,mark=*] 
			coordinates {(-0.037771,-0.0065528) (-0.08807,-0.019901) (-0.13662,-0.039497) (-0.18725,-0.064978) (-0.23673,-0.093634) 
			(0.017856,-0.003952)(0.06415,-0.015716) (0.11825,-0.048011) (0.16522,-0.089897) (0.21626,-0.1461) (0.2655,-0.20811) 
			(0.31431,-0.27527) (0.36526,-0.35008) (0.41808,-0.43164) (0.46763,-0.51111) (0.51715,-0.5929) (0.56592,-0.67534) 
			(0.61923,-0.76728) (0.66426,-0.84619) (0.71426,-0.93495) (0.76532,-1.0267)}; \addlegendentry{Numerical}
		\addplot[smooth,dashed,line width=1,black,mark=.] 
			coordinates {(-0.193,0.1) (-0.193,-1.7)};
		\addplot[smooth,dashed,line width=1,black,mark=.] 
			coordinates {(0.577,0.1) (0.577,-1.7)};
		\node at (axis cs:-0.85,-0.05) {\textbf{(a)}};
	\end{axis} 
\end{tikzpicture}
%\hspace{1cm}
\begin{tikzpicture}
	%\begin{axis}[width=7.2cm,height=7cm, grid=both, xmin=-1,xmax=1,ymin=-1.7,ymax=0.1, minor tick num=4, xlabel=$\chi$,ylabel=$g_{\theta=0}(\chi)$,title={Surface energy correction for $\mathcal{H}_{\theta=0}$}, legend style={at={(0.5,0.02)},anchor=south}]
	\begin{axis}[width=7.2cm,height=7cm, grid=both, xmin=-1,xmax=1,ymin=-1.7,ymax=0.1, minor tick num=4, xlabel=$\chi$]
		\addplot[smooth,blue] 
			coordinates {(-1,-1.6272) (-0.95,-1.5325) (-0.9,-1.4383) (-0.85,-1.3446) (-0.8,-1.2513) (-0.75,-1.1586) 
			(-0.7,-1.0665)(-0.65,-0.9752) (-0.6,-0.88472) (-0.55,-0.79521) (-0.5,-0.70683) (-0.45,-0.61979) (-0.4,-0.53434) (-0.35,-0.4508)
			(-0.3,-0.36961) (-0.25,-0.29134) (-0.2,-0.21681) (-0.15,-0.1472) (-0.1,-0.084404) (-0.05,-0.031919) (0,0) (0.05,-0.027921)
			(0.1,-0.0682) (0.15,-0.11234) (0.2,-0.15838) (0.25,-0.20551) (0.3,-0.25333) (0.35,-0.30161) (0.4,-0.35022) (0.45,-0.39908) 
			(0.5,-0.44812) (0.55,-0.4973) (0.6,-0.5466) (0.65,-0.59598) (0.7,-0.64545) (0.75,-0.69497) (0.8,-0.74454) (0.85,-0.79416) 
			(0.9,-0.84382) (0.95,-0.8935) (1,-0.94322)}; %\addlegendentry{Analytical}		
		\addplot[only marks,red,mark=*] coordinates {(-0.015452,-0.0060141) (-0.063007,-0.043428) (-0.1108,-0.09634) (-0.15898,-0.15832) 
			(-0.20606,-0.22466) (0.033602,-0.015863) (0.080599,-0.051065) (0.12889,-0.092482) (0.17736,-0.13642) (0.22453,-0.18044) 
			(0.27169,-0.22523) (0.32125,-0.27283) (0.36913,-0.3192) (0.41808,-0.36689) (0.46292,-0.41076)(0.51199,-0.45893) 
			(0.56029,-0.50646) (0.60696,-0.55249) (0.65762,-0.60254) (0.70565,-0.65006) (0.74989,-0.69388)}; %\addlegendentry{Numerical}
		\addplot[smooth,dashed,line width=1,mark=.] 
			coordinates {(-0.193,0.1) (-0.193,-1.7)};
		\addplot[smooth,dashed,line width=1,mark=.] 
			coordinates {(0.577,0.1) (0.577,-1.7)};
		\node at (axis cs:-0.85,-0.05) {\textbf{(b)}};
	\end{axis} 
\end{tikzpicture}
\end{center}
\end{figure}

Putting these results to together we are able to determine the ground state
energies of $\mathcal{H}_{\theta=\pi}$ and $\mathcal{H}_{\theta=0}$ up to
order $\mathcal{L}^0$
\begin{equation}
  \label{eqnEnOpen}
  \begin{aligned}
    E_{\theta=\pi}(\chi_{1}^{+},\chi_{1}^{-}) & = -\left[ \frac{1}{\pi} +
      \frac{2\sqrt{3}}{9} \right] \mathcal{L} + \left[ \frac{3}{2} -
      \frac{2\sqrt{3}}{3} \right] + g_{\theta=\pi}(\chi_{1}^{+})  +
    g_{\theta=\pi}(\chi_{1}^{-}) + o(\mathcal{L}^{0})\,, \\ 
    E_{\theta=0}(\chi_{1}^{+},\chi_{1}^{-}) & =
    -\left[\frac{1}{2\pi}-\frac{2\sqrt{3}}{9}+\frac{3}{4}\right]\mathcal{L} +
    \left[-\frac{3}{4} + \frac{2\sqrt{3}}{3} \right]+
    g_{\theta=0}(\chi_{1}^{+})  + g_{\theta=0}(\chi_{1}^{-}) + o(\mathcal{L}^{0})\,.  
  \end{aligned}
\end{equation}
In Figure \ref{figBoundaryCorrection} we plot the predicted functions
$g_{\theta=\pi}(\chi)$ and $g_{\theta=0}(\chi)$ compared to the equivalent
numerical values obtained by solving the the Bethe equations (with
$\chi_{1}^{-} = \chi_{2}^{\pm} = 0$). \\

\noindent
Although the numerical values have been obtained for only 100 sites we clearly
see that values match the analytically predicted results, both inside and
outside the region $-\frac{1}{3\sqrt{3}} < \chi <\frac{1}{\sqrt{3}}$.  The
numerical solution of the Bethe equations for larger boundary amplitudes is
limited by numerical instabilities resulting from, e.g., boundary Bethe
roots passing the other Bethe roots.
The asymptotic behaviour of the surface energy corrections for large $\chi$
does, however, coincide with what is expected from the corresponding
eigenvalues of the boundary operators $B^{(k)\pm}$ (\ref{eqnBfields}), i.e.\
$-1$ and $2$.  This leads us to conjecture that the energies are analytically
correct for all values of $\chi_{k}^{\pm}\in\R$.

Using Equation (\ref{GSEForm}) along with the relations,
$$ 
E_{h}^{(k)} = -E_{\theta=\pi}(\chi_{k}^{+},\chi_{k}^{-}) \hspace{1cm}
\mbox{and} \hspace{1cm} E_{l}^{(k)} =
E_{\theta=0}(\chi_{k}^{+},\chi_{k}^{-}), 
$$ 
we are able to determine the ground state energy for of $\mathcal{H}_{\theta}$
for generic $\theta$. We should again recall that we have the constraint
$\chi_{1}^{+} \chi_{2}^{+} = \chi_{1}^{-} \chi_{2}^{-} = 0$ and that the
interacting boundary terms break the $D(D_{3})$ invariance of the model.

%%%%%%%%%%%%%%%%%%%%%%%%%%%%%%%%%%%%%%%%%%%%%%%%%%%%%%%%%%%%%%%%%%%%%%%%%%%%%%%%
%%%%%%%%%%%%%%%%%%%%%%%%%%%%%%%%%%%%%%%%%%%%%%%%%%%%%%%%%%%%%%%%%%%%%%%%%%%%%%%%
%%%%%%%%%%%%%%%%%%%%%%%%%%%%%%%%%%%%%%%%%%%%%%%%%%%%%%%%%%%%%%%%%%%%%%%%%%%%%%%%

\section{Excitations and Conformal Field Theories}
The presence of a coupling parameter makes the identification of a conformal
field theory a difficult task.  Every energy level of the Hamiltonian depends
on the coupling parameter $\theta$, moreover which of the energies is the
lowest will change with $\theta$.  This implies that the model can not be
described by a single conformal field theory but rather but multiple ones.  To
simplify the issue we first study the model at the level crossings, i.e.\ for
coupling parameters being integer multiples of $\frac{\pi}{2}$.  As discussed
above the energy spectrum at these points is that of the partial Hamiltonians
$\mathcal{H}_{\theta=0}=\mathcal{H}^{(1)}$ and
$\mathcal{H}_{\theta=\pi}=-\mathcal{H}^{(1)}$ (up to degeneracies).  For these
we can use powerful machinery to accurately describe the model at these
special points.  It turns out that the low energy effective theories of the
partial models are given by minimal models.

Once the critical theories at these special couplings have been identified it
is a relatively straight-forward, albeit non-trivial, task to obtain the
conformal operator content of the complete model for general coupling
$\theta$:
the main difficulties come with identifying the previously mentioned pairing
rules and connecting them with some conserved quantity of the model.

To make the presentation self-contained we begin by presenting our results on
the low energy spectrum of the periodic spin chain reported previously
\cite{FFL2011,FinFra2012} withan extended discussion of the residual $D(D_3)$
symmetry under these boundary conditions and the pairing rules.  This section
is followed by new results of our studies of the critical properties of the
periodic fusion path chain and the models with braided and open boundary
conditions.

\subsection{Periodic Spin Chain}
As discussed above the spectrum at the the level crossings is expected to be
described by a single conformal field theory.  As a consequence of conformal
invariance the scaling behaviour of the ground state energy is predicted to be
\cite{BlCN86,Affl86} 
$$ E = \epsilon_{\infty}\L - (c\, v_{F}) \times \frac{\pi}{6\L} + o(\L^{-1}), $$
where $c$ is the central charge of the underlying Virasoro algebra.  For a
given realization of the CFT its operator content is constrained by modular
invariance of the partition function and the particular choice of boundary
conditions \cite{Card86b,CaIZ87}.  Further constraints are imposed by locality
of the physical fields.
The primary fields present in the critical model determine the finite size
energies and partial momenta of the excitated states:
\begin{equation}
\label{cft}
  E(\L) - E_0(\L) = \frac{2\pi v_F}{\L} \left(X+n+\bar{n}\right)\,,\quad
  P(\L) - P_0(\L) = \frac{2\pi}{\L} \left(s+n-\bar{n}\right) + \mbox{const.}\,.
\end{equation}
This allows us to determine the scaling dimensions $X=h+\bar{h}$ and conformal
spins $s = h-\bar{h}$ of the primary fields ($n$, $\bar{n}$ are non-negative
integers) from numerical finite size data obtained by solution to the Bethe
equations along with Equations (\ref{eqnEnCom}) and (\ref{eqnMoCom}).  Note
that due to the massive degeneracies appearing in the spectrum for couplings
$\theta$ being integer multiples of $\pi/2$ the complete momenta are not
unique.  The partial ones entering (\ref{cft}), however, are.  This allows the
use of finite size data at the level crossings for the identification of the
critical theory.

\subsubsection{Spectrum of $\mathcal{H}_{\theta=\pi}$} 
\label{SecSpecPeriodHigh}
The ground state energy of $\mathcal{H}_{\theta=\pi}$ is known to be
\cite{AlDM92,FFL2011}
\begin{equation}
 \label{eqnGSHigh}
 E_{0}  = -\left[ \frac{1}{\pi} + \frac{2\sqrt{3}}{9} \right] \L -
 \frac{12}{5}\times \frac{\pi}{6\L} + {o}(\L^{-1}) \,.
\end{equation}
Using the Fermi-velocity (\ref{eqnvFPI}) computed before the central charge of
the effective field theory for the low energy degrees of freedom in
$\H_{\theta=\pi}$ is identified to be $c=4/5$.  Hence, this sector of the
model is in the universality class of the minimal model $\mathcal{M}_{(5,6)}$
and the conformal weights $h$, $\bar{h}$ of the primary fields can take the
rational values from the Kac table
\begin{eqnarray*}
  h,\bar{h} 
  & \in & \left\{\left.\frac{(6p-5q)^{2}-1}{120}\, \right|\, 1 \leq q
    \leq p < 5 \right\} \\ 
  & = & \left\{ 0, \frac{1}{40}, \frac{1}{15}, \frac{1}{8},
    \frac{2}{5},\frac{21}{40}, \frac{2}{3}, \frac{7}{5}, \frac{13}{8}, 3
  \right\} 
\end{eqnarray*}
To identify the operator content of the periodic spin chain we have solved the
Bethe equations (\ref{eqnBAperiodic}) for lattice sizes up to a
minimum of $\mathcal{L}=40$, although over 100 sites were considered whenever
possible.  The sequence of finite size estimations for the scaling dimensions
\begin{equation*}
  X^{\mathrm{num}}_\theta (\mathcal{L})=\frac{\mathcal{L}}{2\pi v_F}
  \left(E(\mathcal{L }) - E_0(\mathcal{L })\right)
\end{equation*}
has then be extrapolated to get a numerical approximation
$X_\theta^{\mathrm{ext}}$ to the scaling dimension which can then be
identified with a pair $(h,\bar{h})$ of conformal weights from the Kac table.
In Table \ref{tab:ExPerHigh0mod2} we present our data for the low lying
excitations appearing in the $\theta=\pi$ sector of the periodic spin chain
for even chain lengths.
%%%%%%%%%%%%%%%%%%%%%%%%%%%%%%%%%%%%%%%%%%%%%%%%%%%%%%%%%%%%%%%%%%%%%%
\begin{table}[ht]
  \caption{\label{tab:ExPerHigh0mod2}Scaling dimensions $X_\pi$ extrapolated
    from the finite size behaviour of the ground state and low energy
    excitations of $\H_{\theta=\pi}$ (periodic) for even $\mathcal{L}$.
    $(h,\bar{h})$ are the predictions from the $\mathcal{M}_{(5,6)}$ minimal
    model. We have also indicated the $D(D_3)$ sector in which the state
    appears and its conjectured degeneracy.  The operator content of the sector
    $\pi_{\mathfrak{d}}$ is obtained from that of $\pi_{\mathfrak{c}}$ by
    interchanging $h$ and $\bar{h}$.}  
\begin{tabularx}{\textwidth}{CCCCC} \hline \hline
	$D(D_3)$ & $X_{\pi}^{\mathrm{ext.}}$  & $(h,\bar{h})$ & spin & degeneracy \tabularnewline \hline
	$\pi_{\mathfrak{a}} \oplus \pi_{\mathfrak{b}}$
		& 0.000000(1)&  $(0,0)$ & $0$ & $1\times 3^{\frac{\mathcal{L}}{2}-1}$ \tabularnewline
		& 0.801(3) &  $(\frac{2}{5},\frac{2}{5})$ & $0$ & $1\times 3^{\frac{\mathcal{L}}{2}-1}$ \tabularnewline
		& 1.80(1) & $(\frac{2}{5},\frac{7}{5}),(\frac{7}{5},\frac{2}{5})$ & $\pm1$ & $1\times 3^{\frac{\mathcal{L}}{2}-1}$ \tabularnewline \hline
	$\pi_{\mathfrak{c}}$
		& 0.4668(2) &  $(\frac{1}{15},\frac{2}{5})$ & $-\frac{1}{3}$ & $2\times 3^{\frac{\mathcal{L}}{2}-1}$ \tabularnewline
		& 0.666666(1) &  $(\frac{2}{3},0)$ & $\frac{2}{3}$ & $2\times 3^{\frac{\mathcal{L}}{2}-1}$ \tabularnewline \hline
	$\pi_{\mathfrak{e}} \oplus \pi_{\mathfrak{f}}$
		& 0.13334(6) &  $(\frac{1}{15},\frac{1}{15})$ & $0$ & $4\times 3^{\frac{\mathcal{L}}{2}-1}$ \tabularnewline
		& 1.33333(3) &  $(\frac{2}{3},\frac{2}{3})$ & $0$ & $4\times 3^{\frac{\mathcal{L}}{2}-1}$ \tabularnewline \hline \hline
\end{tabularx}
\end{table}
%%%%%%%%%%%%%%%%%%%%%%%%%%%%%%%%%%%%%%%%%%%%%%%%%%%%%%%%%%%%%%%%%%%%%%

\noindent
For the analysis of the finite size spectrum it is convenient to classify
excitations of the model in terms of symmetry sectors.  For even $\mathcal{L}$
the Hilbert space of the spin chain can be decomposed as
$$ 
 \pi_{\mathfrak{g}}^{\tp\mathcal{L}} = \frac{1}{2}(3^{\mathcal{L}-2}+1)
 \pi_{\mathfrak{a}} \oplus \frac{1}{2}(3^{\mathcal{L}-2}-1) \pi_{\mathfrak{b}}
 \oplus 3^{\mathcal{L}-2}\left[\pi_{\mathfrak{c}} \oplus \pi_{\mathfrak{d}}
   \oplus \pi_{\mathfrak{e}} \oplus \pi_{\mathfrak{f}} \right]. 
$$ 
However, as periodic closure of the system breaks the $D(D_{3})$ invariance of
the model we can no longer use this decomposition directly.  Instead we find
it useful define four residual symmetry sectors, i.e.\ $\pi_{\mathfrak{a}}
\oplus \pi_{\mathfrak{b}}$, $\pi_{\mathfrak{c}}$, $\pi_{\mathfrak{d}}$ and
$\pi_{\mathfrak{e}} \oplus \pi_{\mathfrak{f}}$.  Here, the symmetry sector
$\pi_{\mathfrak{a}} \oplus \pi_{\mathfrak{b}}$ is defined to be the subspace
of $\pi_{\mathfrak{g}}^{\tp\mathcal{L}}$ composed of all one-dimensional
irreps ($\pi_{\mathfrak{a}}$ and $\pi_{\mathfrak{b}}$) appearing in its
decomposition, likewise for the other sectors.  Under this definition we find
that all of these (non-intersecting) symmetry sectors are invariant under the
action of global Hamiltonian of the periodic spin chain with generic $\theta$.

Apart from fixing transformation properties of the excited states their
classification according to symmetry counting arguments can be used to
conjecture the degeneracy of each excitation for large finite systems based on
the finite-size spectra, also shown in Table \ref{tab:ExPerHigh0mod2}.
Furthermore, it has been noticed \cite{FinFra2012} that the symmetry of the
eigenstates is connected to the number of Bethe roots at $\pm\infty$, see
Table \ref{tab:SymSectInftyRootsEven}.  Finally, we observe that every
excitation $(h,\bar{h})$ appearing in the sector $\pi_{\mathfrak{c}}$ can be
related to an excitation $(\bar{h}, h)$ appearing in sector
$\pi_{\mathfrak{d}}$ via a mapping of Bethe roots, $\{x_{j}\} \rightarrow
\{-x_{j}\}$.

\begin{table}[ht]
  \caption{\label{tab:SymSectInftyRootsEven}The classification of
    symmetry sectors of the periodic spin chain in terms the numbers of finite
    and infinite Bethe roots for chains of even length.}  
\begin{tabularx}{\textwidth}{CCCC} \hline \hline
	Symmetry Sector & $N_{-}+N_{+}+2N_{s}$ & $n_{-\infty}$  & $n_{+\infty}$ \tabularnewline \hline
	$\pi_{\mathfrak{a}} \oplus \pi_{\mathfrak{b}}$ & $\mathcal{L}$ & 0 & 0 \tabularnewline
	$\pi_{\mathfrak{c}}$ & $\mathcal{L}-1$ & 0 & 1 \tabularnewline
	$\pi_{\mathfrak{d}}$ & $\mathcal{L}-1$ & 1 & 0 \tabularnewline
	$\pi_{\mathfrak{e}} \oplus \pi_{\mathfrak{f}}$ & $\mathcal{L}-2$ & 1 & 1 \tabularnewline \hline \hline
\end{tabularx}
\end{table}

The excitations appearing here for even $\mathcal{L}$ coincide with those from
the self-dual ferromagnetic 3-state Potts quantum chain subjected to either
periodic of twisted boundary conditions \cite{GehRit1986}. This stems from
both models being constructed from the same set of solutions to the
star-triangle equation \cite{FatZam1982b}, albeit with different limits
applied.  The excitations in the $\pi_{\mathfrak{a}} \oplus
\pi_{\mathfrak{b}}$ and $\pi_{\mathfrak{e}} \oplus \pi_{\mathfrak{f}}$ sectors
have been observed in the charge $Q=0$ and $Q=1$ sector of the periodic
3-state Potts chain, respectively.
The excitations in the $\pi_{\mathfrak{c}}$ and $\pi_{\mathfrak{d}}$ sectors
correspond to the 3-state Potts chain with twisted boundary conditions
allowing for the non-half-integer spins observed.  The excitations appearing
in the periodic Potts chain have also been determined using Bethe ansatz
methods by Albertini \emph{et al.} \cite{AlDM92}.

We have also determined the excitations for chains of odd length. In this case
the states can be classified by the eigenvalue of the state under the $D_{3}$
rotation operator, $\sigma$. The excitations are given Table
\ref{tab:ExPerHigh1mod2}.

\begin{table}[ht]
\caption{\label{tab:ExPerHigh1mod2}As Table~\ref{tab:ExPerHigh0mod2} for $\H_{\theta=\pi}$ (periodic) when is $\mathcal{L}$ odd. Symmetry is classified by the action of the $D_3$ rotation $\sigma$.}
\begin{tabularx}{\textwidth}{CCCCC} \hline \hline
	$\sigma$ & $X_{\pi}^{\mathrm{ext.}}$ & $(h,\bar{h})$ & spin & degeneracy \tabularnewline \hline
	1
		& 0.125000(5) & $(0,\frac{1}{8})$ & $-\frac{1}{8}$ & $1\times 3^{\frac{\mathcal{L}-1}{2}}$ \tabularnewline
		& 0.42502(2) & $(\frac{2}{5},\frac{1}{40})$ & $\frac{3}{8}$ & $1\times 3^{\frac{\mathcal{L}-1}{2}}$ \tabularnewline
		& 0.92490(6) & $(\frac{2}{5},\frac{21}{40})$ & $-\frac{1}{8}$ & $1\times 3^{\frac{\mathcal{L}-1}{2}}$ \tabularnewline
		& 1.625000(1) & $(0,\frac{13}{8}) $ & $-\frac{13}{8}$ & $1\times 3^{\frac{\mathcal{L}-1}{2}}$ \tabularnewline \hline
	$\omega$, $\omega^{-1}$
		& 0.091665(2) & $ (\frac{1}{15},\frac{1}{40})$ & $\frac{1}{24}$ & $2\times 3^{\frac{\mathcal{L}-1}{2}}$ \tabularnewline
		& 0.59168(7) & $ (\frac{1}{15},\frac{21}{40})$ & $-\frac{11}{24}$& $2\times 3^{\frac{\mathcal{L}-1}{2}}$ \tabularnewline
		& 0.791667(1) & $ (\frac{2}{3},\frac{1}{8}) $ & $\frac{13}{24}$ & $2\times 3^{\frac{\mathcal{L}-1}{2}}$ \tabularnewline \hline \hline
\end{tabularx}
\end{table}

\noindent
These excitations are not present in the 3-state Potts model since the
equivalence to this model is restricted to $\mathcal{L}$ even.  We were able
to again classify the symmetry sectors in terms of Bethe roots as presented in
Table \ref{tab:SymSectInftyRootsOdd}. We found that $n_{-\infty}=0$ in the
case of odd length chains.

\begin{table}[ht]
  \caption{\label{tab:SymSectInftyRootsOdd}The symmetry sectors of the
    periodic spin chain classified in terms of the numbers of finite and
    infinite Bethe roots for chains of odd length.}  
\begin{tabularx}{\textwidth}{CCCC} \hline \hline
	Symmetry Sector & $N_{-}+N_{+}+2N_{s}$ & $n_{-\infty}$  & $n_{+\infty}$ \tabularnewline \hline
	$1$ & $\mathcal{L}$ & 0 & 0 \tabularnewline
	$\omega,\omega^{-1}$ & $\mathcal{L}-1$ & 0 & 1 \tabularnewline \hline \hline
\end{tabularx}
\end{table}

\subsubsection{Spectrum of $\mathcal{H}_{\theta=0}$}
The ground state energy is known to be \cite{AlDM92,FFL2011},
\begin{equation}
  E_{0} =  -\left[\frac{1}{2\pi} - \frac{2\sqrt{3}}{9} +
    \frac{3}{4}\right] \mathcal{L} - \frac{3}{2}\times
  \frac{\pi}{6\mathcal{L}} + {o}(\mathcal{L}^{-1}) \label{eqnGSLow} 
\end{equation}
Using the Fermi-velocity we find that the central charge is $1$, which does
not uniquely define a conformal field theory.
The field content of the theory is obtained from the finite size spectrum.
One method of determining the the finite size spectrum is using the dressed
charge formalism (see appendix) leading to the identification of the $Z_{4}$
parafermion theory \cite{ZaFa85,GeQi87} coinciding with the anti-ferromagnetic
3-state Potts model. The allowed conformal weights for this theory are
\cite{GeQi87,CoyKed1993}
\begin{eqnarray*}
  h,\bar{h} 
  & \in & \left\{ \left.\frac{l(l+2)}{24} - \frac{m^{2}}{16} \right| 0 \leq m \leq l \leq 4,\, l\equiv m\, (\mbox{mod}\, 2) \right\} \\ \\
  & = & \left\{ 0, \frac{1}{16}, \frac{1}{12}, \frac{1}{3}, \frac{9}{16}, \frac{3}{4}, 1 \right\}
\end{eqnarray*}
Alternately we can solve the Bethe equations directly and determine the
scaling behaviour of the low-lying excitations. Here we must consider
$\mathcal{L}=0,1,2,3 \pmod{4}$ separately, see Tables
\ref{tab:ExPerLow0mod4}-\ref{tab:ExPerLow2mod4} below.  In particular, we find
that the finite size gap of the lowest states for
$\ell=\mathcal{L}\,(\mbox{mod}\,4)\ne0$ is determined by an (anti-)chiral
$Z_{k=4}$ spin field with conformal weight $h_\ell= \ell(k-\ell)/(2k(k+2))$.

\begin{table}[ht]
	\caption{\label{tab:ExPerLow0mod4}Scaling dimensions $X_0$
          extrapolated from the finite size behaviour of the ground state and
          low energy excitations of $\mathcal{H}_{\theta=0}$ (periodic) for
          $\mathcal{L}=0\pmod{4}$ (the error of the extrapolation is smaller
          than the last displayed digit).  $(h,\bar{h})$ are the predictions
          from the $Z_4$ parafermionic CFT.  For the other columns, see
          Table~\ref{tab:ExPerHigh0mod2}.} 
\begin{tabularx}{\textwidth}{CCCCC} \hline \hline
	$D(D_3)$ & $X_{0}^{\mathrm{ext.}}$ & $(h,\bar{h})$ & spin & degeneracy\tabularnewline \hline
	$\pi_{\mathfrak{a}} \oplus \pi_{\mathfrak{b}}$ 
		& $0.000000$ & $(0,0)$ & $0$ & $1\times 3^{\frac{\mathcal{L}}{2}-1}$ \tabularnewline \hline 
	$\pi_{\mathfrak{c}}$
		& $0.333332$ & $(0,\frac{1}{3})$ & $-\frac{1}{3}$ & $2\times 3^{\frac{\mathcal{L}}{2}-1}$ \tabularnewline \hline 
	$\pi_{\mathfrak{e}} \oplus \pi_{\mathfrak{f}}$ 
		& $0.166667$ & $(\frac{1}{12},\frac{1}{12})$ & $0$ & $4\times 3^{\frac{\mathcal{L}}{2}-1}$ \tabularnewline
		& $0.666667$ & $(\frac{1}{3},\frac{1}{3})$ & $0$ & $4\times 3^{\frac{\mathcal{L}}{2}-1}$ \tabularnewline \hline \hline
\end{tabularx}
\end{table}

\begin{table}[ht]
  \caption{\label{tab:ExPerLow1mod4}As Table \ref{tab:ExPerLow0mod4} for
    $\mathcal{H}_{\theta=0}$ (periodic) when $\mathcal{L}=1\pmod{4}$. The
    excitations for chains with length $\mathcal{L}=3\pmod{4}$ have the same
    exponents but the opposite spin.  Symmetry is classified by the action of
    the $D_3$ rotation $\sigma$.} 
\begin{tabularx}{\textwidth}{CCCCC} \hline \hline
     $\sigma$ & $X_{0}^{\mathrm{ext.}}$ & $(h,\bar{h})$ & spin & degeneracy \tabularnewline \hline \hline
     $1$ 
     & $0.062500$ & $(\frac{1}{16},0)$ & $\frac{1}{16}$ & $1\times 3^{\frac{\mathcal{L}-1}{2}}$ \tabularnewline 
     & $0.562500$ & $(\frac{9}{16},0)$ & $\frac{9}{16}$ & $1\times 3^{\frac{\mathcal{L}-1}{2}}$ \tabularnewline 
     & $0.812500$ & $(\frac{1}{16},\frac{3}{4})$ & $-\frac{11}{16}$ & $1\times 3^{\frac{\mathcal{L}-1}{2}}$ \tabularnewline \hline 
     $\omega,\omega^{-1}$
     & $0.145833$ & $(\frac{1}{16},\frac{1}{12})$ & $-\frac{1}{48}$ & $2\times 3^{\frac{\mathcal{L}-1}{2}}$ \tabularnewline
     & $0.395833$ & $(\frac{1}{16},\frac{1}{3})$ & $-\frac{13}{48}$ & $2\times 3^{\frac{\mathcal{L}-1}{2}}$ \tabularnewline
     & $0.645833$ & $(\frac{9}{16},\frac{1}{12})$ & $\frac{23}{48}$ & $2\times 3^{\frac{\mathcal{L}-1}{2}}$ \tabularnewline \hline \hline
\end{tabularx}
\end{table}

\begin{table}[ht]
	\caption{\label{tab:ExPerLow2mod4}As Table \ref{tab:ExPerLow0mod4} but
          for $\mathcal{H}_{\theta=0}$ (periodic) when
          $\mathcal{L}=2\pmod{4}$.} 
\begin{tabularx}{\textwidth}{CCCCC} \hline \hline
	$D(D_3)$ & $X_{0}^{\mathrm{ext.}}$ & $(h,\bar{h})$ & spin & degeneracy \tabularnewline  \hline \hline
	$\pi_{\mathfrak{a}} \oplus \pi_{\mathfrak{b}}$
		& $0.750000$ & $(0,\frac{3}{4})\times2,$ & $\pm\frac{3}{4}$ & $1\times 3^{\frac{\mathcal{L}}{2}-1}$ \tabularnewline
		& & $(\frac{3}{4},0)\times2$ & & \tabularnewline\hline  
	$\pi_{\mathfrak{c}}$
		& $0.083333$ & $(0,\frac{1}{12})$ & $-\frac{1}{12}$ & $2\times 3^{\frac{\mathcal{L}}{2}-1}$ \tabularnewline 
		& $1.083333$ & $(\frac{3}{4},\frac{1}{3})$ & $\frac{5}{12}$ & $2\times 3^{\frac{\mathcal{L}}{2}-1}$ \tabularnewline \hline 
	$\pi_{\mathfrak{e}} \oplus \pi_{\mathfrak{f}}$
		& $0.416667$ & $(\frac{1}{12},\frac{1}{3}),(\frac{1}{3},\frac{1}{12})$ & $\pm \frac{1}{4}$ & $4\times 3^{\frac{\mathcal{L}}{2}-1}$\tabularnewline \hline \hline
\end{tabularx}
\end{table}

\noindent
As with the previous case we again can partition the excitations according to
the residual symmetry sectors. These sectors are still characterised by the
number of Bethe roots at $\pm\infty$ as described in Table
\ref{tab:SymSectInftyRootsEven}.

Comparing these excitations to the anti-ferromagnetic 3-state Potts chain
\cite{AlDM92,CoyKed1993} (for $\mathcal{L}$ even), we find that the
excitations in the $\pi_{\mathfrak{a}} \oplus \pi_{\mathfrak{b}}$ and
$\pi_{\mathfrak{e}} \oplus \pi_{\mathfrak{f}}$ were previously identified and
restricted to the $n_{+\infty}=n_{-\infty}$ case.  We were unable to find any
literature dealing with the anti-ferromagnetic end of the twisted 3-Potts
model.  We expect, however, that the excitations in that case to match those
appearing in the $\pi_{\mathfrak{c}}$ and $\pi_{\mathfrak{d}}$ sectors.
Similarly the excitations for odd $\mathcal{L}$ have not previously been
studied.

\subsubsection{Pairing rules and discussion}
The results on the low energy spectra $\mathcal{H}_\theta$ for $\theta=0,\pi$
completely determine the critical behaviour (up to degeneracies) of the
periodic $D(D_3)$ spin chain at all the level crossings, i.e.\ when $\theta$
is a multiple of $\frac{\pi}{2}$: this is a consequence of the fact that the
partial energies, i.e.\ eigenvalues (\ref{eqnEnCom}) of $\mathcal{H}^{(1)}$
and $\mathcal{H}^{(2)}$, and momenta (\ref{eqnMoCom}) corresponding to a given
Bethe root configuration are identical (although the partial momenta enter in
the definition of the total momentum (\ref{eqnMotot}) with opposite signs).

For generic values of $\theta$ the energy eigenvalues of the periodic spin
chain are given by (\ref{eqnEtot}) in terms of two root configurations of the
Bethe equations (\ref{eqnBAperiodic}) provided that these configurations pair.
Based on studies of small system sizes it has been observed that two sets of
Bethe roots pair to form an eigenvalue of the transfer matrix, Equation
(\ref{eqnTranferEig}), if and only if they have matching number of roots at
$\pm\infty$ \cite{FFL2011}.  As a consequence of the classification of
excitations according to their symmetry above, see Tables
\ref{tab:SymSectInftyRootsEven} and \ref{tab:SymSectInftyRootsOdd}, 
this is equivalent to saying Bethe root configurations (and their
corresponding energies) pair if and only if they belong to the same symmetry
sector.  Furthermore, we have observed that within a symmetry sector pairing
is uniform in the sense that every two sets of Bethe root configurations
within a symmetry sector pair the same number of times, a quantity referred to
as the pairing multiplicity. The relationship between pairing multiplicity and
symmetry sectors is documented in Table \ref{tab:PairingMultiplicity}.

\begin{table}[ht]
  \caption{\label{tab:PairingMultiplicity}The pairing multiplicities, $m_{p}$,
    of the periodic spin chain for any two energies of $\mathcal{H}^{(1)}$ and
    $\mathcal{H}^{(2)}$ that belong to the same symmetry sector. If the
    energies do not belong to the same symmetry sector then they do not pair,
    i.e. the pairing multiplicity is zero.} 
\begin{tabularx}{\textwidth}{CCCCCCC} \hline \hline
	& \multicolumn{4}{c}{$\mathcal{L}=0\pmod{2}$} & \multicolumn{2}{c}{$\mathcal{L}=1\pmod{2}$}  \tabularnewline \hline \hline
	Sector & $\pi_{\mathfrak{a}} \oplus \pi_{\mathfrak{b}}$  & $\pi_{\mathfrak{c}}$ & $\pi_{\mathfrak{d}}$ & $\pi_{\mathfrak{e}} \oplus \pi_{\mathfrak{f}}$ & 1 & $w,w^{-1}$  \tabularnewline
	$m_{p}$ & 1 & 2 & 2 & 4 & 1 & 2 \tabularnewline \hline \hline
\end{tabularx}
\end{table}

\noindent
This information, along with Equations (\ref{eqnEtot}) and (\ref{cft}) and the
relevant tables, is sufficient to determine the energies and degeneracies of
the ground state and low lying excitations of the model for generic $\theta$.
The resulting spectrum is that of a direct product of two conformal field
theories.  The physical fields appearing in the combined theory are composite
operators with scaling dimension $X_{\mathrm{tot}}=X^{(1)}+X^{(2)} =
\sum_{k=1}^2 (h^{(k)}+\bar{h}^{(k)})$ with $(h^{(k)},\bar{h}^{(k)})$ being the
conformal weights from the two components.
Similarly, the total spin of an excitation can be calculated from Equations
(\ref{eqnMotot}) and (\ref{cft}) giving $s_{\mathrm{tot}}=s^{(1)}-s^{(2)} =
h^{(1)}-\bar{h}^{(1)} -h^{(2)}+\bar{h}^{(2)} $.
Note that application of the pairing rules to the conformal dimensions
identified for the periodic spin chain above ensure that this total spin is
always either an integer or half-integer which guarantees locality of the
physical fields.

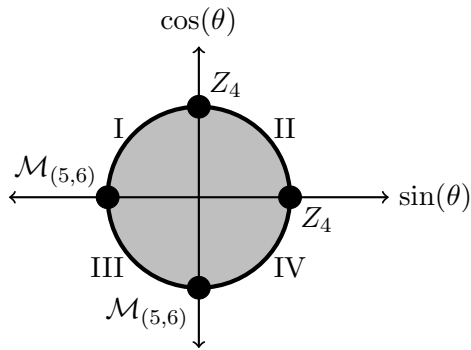
\begin{figure}[ht]
  \caption{Conformal field theory description of the periodic integrable
    $D(D_{3})$ symmetric model for generic $\theta$. Level crossing are
    described by minimal models, either $Z_{4}$ parafermions or
    $\mathcal{M}_{(5,6)}$, while the regions in between are described by a
    product of the theories of the two adjacent level crossings. Therefore
    regions I and IV correspond to a $Z_{4}\tp \mathcal{M}_{(5,6)}$ theory,
    while regions II and III correspond respectively to $Z_{4}\tp Z_{4}$ and
    $\mathcal{M}_{(5,6)}\tp \mathcal{M}_{(5,6)}$
    theories. \label{FigCFTgenericTheta}}
\begin{center}
\begin{tikzpicture}[scale=1.0]
	%Basic graph structure
	\draw [black, ultra thick,fill=lightgray] (0,0) circle [radius=1.2];
	\draw [thick,<->] (2.5,0) -- (-2.5,0);
	\draw [thick,<->] (0,2) -- (0,-2);	
	\node [above] at (0,2.0) {$\cos(\theta)$};
	\node [right] at (2.5,0) {$\sin(\theta)$};
	
	\draw [fill] (0,1.2) circle [radius=0.15];
	\draw [fill] (0,-1.2) circle [radius=0.15];
	\draw [fill] (1.2,0) circle [radius=0.15];
	\draw [fill] (-1.2,0) circle [radius=0.15];

	\node [above right] at (0,1.2) {$Z_{4}$};
	\node [below left] at (0,-1.2) {$\mathcal{M}_{(5,6)}$};
	\node [below right] at (1.2,0) {$Z_{4}$};
	\node [above left] at (-1.2,0) {$\mathcal{M}_{(5,6)}$};
	
	\node [right] at (0.85,0.92) {II};
	\node [left] at (-0.85,0.92) {I};
	\node [right] at (0.85,-0.92) {IV};
	\node [left] at (-0.85,-0.92) {III};
\end{tikzpicture}
\end{center}
\end{figure}

The phase diagram of the complete model can be summarised by Figure
\ref{FigCFTgenericTheta}.

\subsection{Periodic Fusion Path Chain}
By construction this chain differs from the periodic spin chain discussed in
the previous section only through boundary conditions.  Therefore, the two
models share their bulk properties, including energy per unit lattice site,
Fermi velocity and central charge.
To identify the operator content of the low energy effective theory for the
fusion path model we have computed the complete spectrum of the Hamiltonian
numerically for up to $\mathcal{L}=10$ sites.  We find that part of the
spectrum coincides (numerically exact) with eigenvalues of the periodic spin
chain (although the corresponding degeneracies in the different formalisms do
not match).  Specifically, this applies to the energies which have been
associated to the $\pi_{\mathfrak{a}}\ds \pi_{\mathfrak{b}}$ and
$\pi_{\mathfrak{e}}\ds \pi_{\mathfrak{f}}$ symmetry sectors above (we
emphasize that the periodic fusion path model can only be constructed for
$\mathcal{L}$ even).  In addition we have diagonalized the transfer matrix for
up to $\mathcal{L}=8$ sites.  From these results we find that the transfer
matrix eigenvalues factorize into two polynomials as in (\ref{eqnTranferEig}).
The corresponding roots of these polynomials can be used to parametrize the
eigenvalues and are conjectured to be given by the Bethe equations
(\ref{eqnBAFusPath}).

We have verified this conjecture by comparing the energies obtained from the
Bethe equations with those obtained by numerical diagonalization.  The
eigenvalues appearing in both the spin chain and the fusion path formalism are
described by root configurations solving (\ref{eqnBAFusPath}) with $\eta=+1$.
This includes the ground states of $\mathcal{H}_{\theta=\pi}$ and
$\mathcal{H}_{\theta=0}$ with energies given by Equations (\ref{eqnGSHigh})
and (\ref{eqnGSLow}), respectively.  Generally, we find that each set of Bethe
roots corresponding to an eigenvalue of the periodic fusion path chain
contains either only finite roots or, for $\eta=+1$, exactly two roots at
$+\infty$ and $-\infty$.

\subsubsection{Spectrum of $\tilde{\mathcal{H}}_{\theta=\pi}$}
As this model shares bulk properties with the periodic spin chain this model
lies in the $M_{(5,6)}$ universality class with central charge
$c=\frac{4}{5}$. 
From the diagonalization of the transfer matrix we can associate root
configurations to the ground state and low lying excitations of
$\tilde{\mathcal{H}}_{\theta=\pi}$ which consist of $\approx \frac{1}{2}
\mathcal{L}$ 2-strings and a few $\pm$-strings.  Solving the conjectured Bethe
equations for systems of up to 100 sites we have found the excitations
given in Table \ref{tab:ExAnyonHigh0mod2}.
\begin{table}[ht]
\caption{\label{tab:ExAnyonHigh0mod2}Scaling dimensions $X_\pi$ extrapolated
  from the finite size behaviour of the ground state and low energy
  excitations of $\tilde{\mathcal{H}}_{\theta=\pi}$ (fusion path) for
  $\mathcal{L}=0\pmod{2}$.  
  From the numerical diagonalisation of the Hamiltonian for finite size
  systems we conjecture the degeneracy of each pair of conformal weights,
  $(h,\bar{h})$. The value $\eta$ is from the Bethe Equations
  (\ref{eqnBAFusPath}). } 
\begin{tabularx}{\textwidth}{CCCCC} \hline \hline
	$X_{\pi}^{\mathrm{ext.}}$ & $(h,\bar{h})$ & spin & degeneracy & $\eta$ \tabularnewline \hline
	$0.000000(1)$ & $(0,0)$ & $0$ & $\frac{1}{2}(3^{\frac{\mathcal{L}}{2}}+1)$ & $+1$ \tabularnewline
	$0.05004(5)$ & $(\frac{1}{40},\frac{1}{40})$ & $0$ & $\frac{1}{2}(3^{\frac{\mathcal{L}}{2}}+1)$ & $-1$ \tabularnewline
	$0.13334(4)$ & $(\frac{1}{15},\frac{1}{15})$ & $0$ & $3^{\frac{\mathcal{L}}{2}}$ & $+1$ \tabularnewline
	$0.25001(3)$ & $(\frac{1}{8},\frac{1}{8})$ & $0$ & $\frac{1}{2}(3^{\frac{\mathcal{L}}{2}}+1)$ & $-1$ \tabularnewline
	$0.549(2)$ & $(\frac{1}{40},\frac{21}{40}),(\frac{21}{40},\frac{1}{40})$ & $\pm\frac{1}{2}$ & $\frac{1}{2}(3^{\frac{\mathcal{L}}{2}}-1)$ & $-1$ \tabularnewline
	$0.801(3)$ & $(\frac{2}{5},\frac{2}{5})$ & $0$ & $\frac{1}{2}(3^{\frac{\mathcal{L}}{2}}+1)$ & $+1$ \tabularnewline
	$1.06(2)$ & $(\frac{21}{40},\frac{21}{40})$ & $0$ & $\frac{1}{2}(3^{\frac{\mathcal{L}}{2}}+1)$ & $-1$ \tabularnewline
	$1.33333(3)$ & $(\frac{2}{3},\frac{2}{3})$ & $0$ & $3^{\frac{\mathcal{L}}{2}}$ & $+1$ \tabularnewline
	$1.80(1)$ & $(\frac{2}{5},\frac{7}{5}),(\frac{7}{5},\frac{2}{5})$ & $\pm 1$ & $\frac{1}{2}(3^{\frac{\mathcal{L}}{2}}-1)$ & $+1$ \tabularnewline \hline \hline
\end{tabularx}
\end{table}

\noindent
Note that the new excitations corresponding to roots of the Bethe equations
(\ref{eqnBAFusPath}) with $\eta=-1$ do not correspond to excitations of the
3-state Potts chains subject to the boundary conditions studied previously
\cite{GehRit1986}.  Moreover, because the formulation of this chain is reliant
on $D(D_{3})$ symmetry, which is not present in the usual 3-state Potts local
Hamiltonian, it is reasonable to expect that these excitations won't appear
for any other formulation of the 3-state Potts model.

\subsubsection{Spectrum of $\tilde{\mathcal{H}}_{\theta=0}$}
Using similar methods we can determine the low energy excitations of the
fusion path model for $\theta=0$.  This is done for even $\mathcal{L}$ where,
as in the spin chain case, we have to discuss the cases of
$\frac{\mathcal{L}}{2}$ even or odd separately.  The conformal dimensions
identified from the low-lying excitations are shown in
Table~\ref{tab:ExAnyonLow}.
\begin{table}[ht]
\caption{\label{tab:ExAnyonLow}Scaling dimensions $X_{0}$ extrapolated from
  the finite size behaviour of the ground state and low energy excitations of
  $\tilde{\mathcal{H}}_{\theta=0}$ (fusion path) for
  $\mathcal{L}=0,2\pmod{4}$. We have omitted the column regarding the
  degeneracy of the excitation due to a lack of data. The error of the
  extrapolation is smaller than the last displayed digit. The association of a
  $\times2$ with a pair of conformal weights implies that there are two
  excitations with those conformal weights that have distinct energies for
  finite size systems.} 
\begin{tabularx}{\textwidth}{CCCCC} \hline \hline
	$\mathcal{L}$ mod 4 & $X_{0}^{\mathrm{ext.}}$ & $(h,\bar{h})$ & spin & $\eta$ \tabularnewline \hline
	0	& $0.000000$ & $(0,0)$ & $0$ & $+1$ \tabularnewline
		& $0.125000$ & $(\frac{1}{16},\frac{1}{16})$ & $0$ & $-1$ \tabularnewline
		& $0.166667$ & $(\frac{1}{12},\frac{1}{12})$ & $0$ & $+1$ \tabularnewline
		& $0.625000$ & $(\frac{9}{16},\frac{1}{16})\times 2,$ & $\pm\frac{1}{2}$ & $-1$ \tabularnewline
		& & $(\frac{1}{16},\frac{9}{16}) \times 2$ &  &  \tabularnewline
		& $0.666667$ & $(\frac{1}{3},\frac{1}{3})$ & $0$ & $+1$ \tabularnewline \hline
	2	& $0.125000$ & $(\frac{1}{16},\frac{1}{16})\times 2$ & $0$ & $-1$ \tabularnewline
		& $0.416667$ & $(\frac{1}{3},\frac{1}{12}),(\frac{1}{12},\frac{1}{3})$ & $\pm\frac{1}{4}$ & $+1$ \tabularnewline
		& $0.625000$ & $(\frac{9}{16},\frac{1}{16})\times 2,$ & $\pm\frac{1}{2}$ & $-1$ \tabularnewline
		& & $(\frac{1}{16},\frac{9}{16}) \times 2$ & & \tabularnewline
		& $0.750000$ & $(0,\frac{3}{4})\times2,$ & $\pm\frac{3}{4}$ & $+1$ \tabularnewline 
		& & $(\frac{3}{4},0)\times2$ & & \tabularnewline \hline \hline
\end{tabularx}
\end{table}

\noindent
As was the case with the $\tilde{\mathcal{H}}_{\theta=\pi}$ model we find new
excitations, again characterised by $\eta=-1$, which do not appear in 3-state
Potts chains subject to the boundary conditions studied previously.
Due to commensurability conditions the spectra can only be compared for
lattices with lengths differing by multiples of 4.  Therefore we do not have
sufficient numerical data from exact diagonalization of the Hamiltonian to
present any conjectures concerning the degeneracy of the excitations.

\subsubsection{Pairing rules and discussion}
In contrast to the periodic spin formulation the pairing rules for the
periodic fusion path chain could not be easily determined.  The residual
symmetry sectors used in the previous section are no longer present.  It is
possible, however, to define a conserved topological charge for the fusion
path model based of the $D(D_{3})$ F-moves aforementioned
\cite{FTLTKWF2007,GATHLTW2013}.  This charge allows to differentiate between
different topological sectors which are labelled by the irreps of $D(D_{3})$.
A better understanding of these topological symmetry sectors will be necessary
to gain insight into the pairing rules for this model.

Finally we want to stress that while the computation of scaling dimensions is
based on the solution of the Bethe equations (\ref{eqnBAFusPath}), which can
be achieved for relatively large $\mathcal{L}$, the identification of allowed
Bethe root configurations still relied upon the explicit diagonalisation of
the transfer matrices of small systems.  Thus while we have a high level of
confidence in the accuracy of the scaling dimension it is not clear whether
all primary excitations have been identified.  In particular, the existence of
additional primary operators with larger scaling dimensions cannot be ruled out.

\subsection{Braided Chain}
As discussed above, the $D(D_3)$ quantum chains in the spin chain and the
fusion path formalism are equivalent for braided and open boundary conditions.
Therefore we discuss of their critical properties in the former. 

\subsubsection{Spectrum of $\mathcal{H}_{\theta=\pi}$}
The full spectrum of the $\mathcal{H}_{\theta=\pi}$ braided chain is a subset
of the spectrum of the $\mathcal{H}_{\theta=\pi}$ periodic spin chain.  In
particular we find that, for the choice of the braiding operator used in this
work, the energies present are those that appeared in the symmetry sectors
$\pi_{\mathfrak{a}} \oplus \pi_{\mathfrak{b}}$ and $\pi_{\mathfrak{c}}$ of the
periodic chain for even $\mathcal{L}$ and in the $\sigma =1$ for odd
$\mathcal{L}$.
As a consequence, only Bethe root configurations with $n_{-\infty}=0$ for even
length chains ($n_{\pm\infty}=0$ for odd length chains) correspond to
eigenvalues of the braided model.  Therefore the low-lying excitations for the
braided chain can be deduced from Tables \ref{tab:ExPerHigh0mod2} and
\ref{tab:ExPerHigh1mod2}:
\begin{eqnarray*}
  (h,\bar{h}) & \in & \left\{\begin{array}{ccc}
      \left\{(0,0),(\frac{2}{5},\frac{2}{5}),(\frac{2}{5},
        \frac{7}{5}),(\frac{7}{5},\frac{2}{5}),(\frac{1}{15},
        \frac{2}{5}),(\frac{2}{3},0)   
      \right\}, && \mathcal{L} \quad \mbox{even}, \\
      \left\{(0,\frac{1}{8}),(\frac{2}{5},\frac{1}{40}),
        (\frac{2}{5},\frac{21}{40}),(0,\frac{13}{8}) 
      \right\}, && \mathcal{L} \quad\mbox{odd}.\end{array}  \right. 
\end{eqnarray*}
Since the braided chain has the full $D(D_{3})$ symmetry it would be possible
to classify all eigenstates of this model in terms of the irreps.  We find,
however, that the enlarged symmetry gives rise to additional degeneracies in
the spectrum which extend across multiple symmetry sectors of the braided spin
chain.
This implies that for these boundary conditions a classification of symmetry
sectors based on the presence of infinite Bethe roots is no longer possible.
In particular, this leads to higher degeneracies in the spectrum of the
braided spin chain model as compared to the periodic one.

\subsubsection{Spectrum of $\mathcal{H}_{\theta=0}$}
As $\mathcal{H}_{\theta=0}=-\mathcal{H}_{\theta=\pi}$ we can again use our
results for the periodic spin chain to discuss the low energy spectrum of the
braided one.  Just as for $\theta=\pi$ the excitations appearing in the
braided chain are those present in the $\pi_{\mathfrak{a}} \oplus
\pi_{\mathfrak{b}}$ and $\pi_{\mathfrak{c}}$ sectors of the periodic chain for
even $\mathcal{L}$ and in the $\sigma =1$ sector for odd $\mathcal{L}$.
Thus we can deduce the excitations appearing in the braided model from Tables
\ref{tab:ExPerLow0mod4}-\ref{tab:ExPerLow2mod4}:
$$ 
(h,\bar{h}) \in \left\{\begin{array}{ccc}
    \left\{(0,0),(0,\frac{1}{3})\right\}, && \mathcal{L} \equiv 0 \mod 4,\\ 
    \left\{(\frac{1}{16},0),(\frac{9}{16},0),(\frac{1}{16},\frac{3}{4})\right\},
    && \mathcal{L} \equiv 1 \mod 4, \\ 
    \left\{(0,\frac{3}{4}),(\frac{3}{4},0), (0,\frac{1}{12}),
      (\frac{3}{4},\frac{1}{3})\right\}, && \mathcal{L} \equiv 2 \mod 4, \\ 
    \left\{(\frac{1}{16},0),(\frac{9}{16},0),(\frac{1}{16},\frac{3}{4})\right\},
    && \mathcal{L} \equiv 3 \mod 4. \\ 
  \end{array}  \right. 
$$

\subsubsection{Pairing rules and discussion}
Like the periodic case the results for the braided chain at $\theta=0,\pi$
determine the low energy spectrum at all of the level crossings. 
Again, the full spectrum for generic $\theta$ is given by (\ref{eqnEtot}) in
terms of two Bethe root configurations provided that these configurations
pair.  For the braided model it has been observed previously, that any two
solutions to the Bethe equations can be paired to form an eigenvalue of the
transfer matrix given by Equation (\ref{eqnTranferEig}) \cite{FFL2011}.  This
is consistent with statement above, that a given root configuration
corresponds to eigenstates in multiple symmetry sectors.  Another difference
to the periodic spin chain is that the assignment of a root configuration to a
particular symmetry sector of the braided chain is different for
$\mathcal{H}^{(1)}$ or $\mathcal{H}^{(2)}$.
Despite these technical differences we find that the pairing multiplicities,
i.e.\ the number of times two solutions to the Bethe equations pair, depends
solely on the symmetry sector the eigenvalue of the transfer matrix lies in
and is defined by Table \ref{tab:PairingMultiplicity}.

As before, pairing determines the physical fields appearing in the low energy
effective theory of the braided spin chain.  Their scaling dimensions and
total spin of these composite operators are related to the ones of their
components as before.  We note, however, that the total momentum in the
braided model is not a multiple of $2\pi/\mathcal{L}$ but instead constrained
by Eq.~(\ref{eqnOrderB}).

\subsection{Open Chain}
As in the case of braided boundary conditions it is sufficient to discuss the
low energy behaviour of the open chain in the spin chain formulation.  At
level crossings conformal invariance predicts that the lowest energies of the
open chain with free ends, i.e. $\chi_{1}^{\pm}=\chi_{2}^{\pm}=0$, have the
scaling behaviour given by \cite{BlCN86}
$$
E = \epsilon_{\infty}\mathcal{L} + f_{0} + \frac{\pi
  v_{F}}{\mathcal{L}}\left(-\frac{c}{24}+h+n\right) + o(\mathcal{L}^{-1}), 
$$ 
where $c$ is the central charge, $v_{F}$ is the Fermi-velocity, $h$ is a
conformal weight and $n$ is a non-negative integer.

\subsubsection{Spectrum of $\mathcal{H}_{\theta=\pi}$}
The ground state energy for $\theta=\pi$ \cite{FFL2011} is,
\begin{eqnarray*}
	E_{0} & = & -\left[ \frac{1}{\pi} + \frac{2\sqrt{3}}{9} \right]
        \mathcal{L} + \left[ \frac{3}{2} - \frac{2\sqrt{3}}{3} \right] -
        \frac{12}{5} \times \frac{\pi}{24\mathcal{L}} + o(\mathcal{L}^{-1}). 
\end{eqnarray*}
As is the case with the periodic chain we have that $v_{F}=3$ and
$c=\frac{4}{5}$, yielding the same CFT as expected. Using an analogous method
to that outlined in Section \ref{SecSpecPeriodHigh} to calculate
$X^{\mathrm{ext.}}$ we can extrapolate values for the conformal weights,
$h^{\mathrm{ext.}}$. The values are presented in Table
\ref{tab:OpenExcitHigh}.

\begin{table}[ht]
  \caption{\label{tab:OpenExcitHigh}Conformal weights $h$ extrapolated from
    the finite size behaviour of the ground state and low energy excitations
    of $\H_{\theta=\pi}$ (open) for different chain lengths. The expression
    for the degeneracy of the energy is written for easy comparison to Table
    \ref{tab:ExPerHigh0mod2}.} 
\begin{tabularx}{\textwidth}{CCCC} \hline \hline
	$\mathcal{L}$ mod 2 & $h^{\mathrm{ext.}}$ & $h$ & degeneracy \tabularnewline \hline \hline
	0 & 0.000000(1) & $0$ & $3\times 3^{\frac{\L}{2}-1}$ \tabularnewline
	  & $0.666(1)$ & $\frac{2}{3}$ & $6\times 3 ^{\frac{\L}{2}-1}$ \tabularnewline \hline
	1 & $0.125(2)$ & $\frac{1}{8}$ & $3\times 3^{\frac{\L-1}{2}}$ \tabularnewline
	  & $1.624(3)$ & $\frac{13}{8}$ & $3\times 3^{\frac{\L-1}{2}}$ \tabularnewline \hline
\end{tabularx}
\end{table}

\subsubsection{Spectrum of $\mathcal{H}_{\theta=0}$}
The ground state of the open chain for $\theta=0$ \cite{FFL2011} is,
\begin{eqnarray*}
	E_{0} & = & -\left[\frac{1}{2\pi} - \frac{2\sqrt{3}}{9} +
          \frac{3}{4}\right] \mathcal{L} + \left[-\frac{3}{4} +
          \frac{2\sqrt{3}}{3} \right] - \frac{3}{2}\times
        \frac{\pi}{24\mathcal{L}} + {o}(\mathcal{L}^{-1}), 
\end{eqnarray*}
which gives $v_{F}=\frac{3}{2}$ and $c=1$, in agreement with the periodic
chain. As with the previous section we numerical approximate the conformal
weights by solving the Bethe equations. The results are summarised in Table
\ref{tab:OpenExcitLow}.
\begin{table}[ht]
  \caption{\label{tab:OpenExcitLow}Conformal weights $h$ extrapolated from the
    finite size behaviour of the ground state and low energy excitations of
    $\H_{\theta=0}$  (open) for different chain lengths. The conformal weights
    appearing when $\mathcal{L}=3$ (mod 4)   have been omitted as they are
    identical to the $\mathcal{L}=1$ (mod 4) case. As was the case with Table
    \ref{tab:ExPerLow0mod4} the error of the extrapolation is smaller than the
    last displayed digit.} 
\begin{tabularx}{\textwidth}{CCCC} \hline \hline
	$\mathcal{L}$ mod 4 & $h^{\mathrm{ext.}}$ & $h$ & degeneracy \tabularnewline \hline \hline
	0 & $0.000000$ & $0$ & $3\times 3^{\frac{\L}{2}-1}$ \tabularnewline
	  & $0.333333$ & $\frac{1}{3}$ & $6\times 3^{\frac{\L}{2}-1}$ \tabularnewline 
	  & $1.000000$ & $1$ & $3\times 3^{\frac{\L}{2}-1}$ \tabularnewline \hline
	1 & $0.062500$ & $\frac{1}{16}$ & $3\times 3^{\frac{\L-1}{2}}$ \tabularnewline
	  & $0.562500$ & $\frac{9}{16}$ & $3\times 3^{\frac{\L-1}{2}}$ \tabularnewline \hline
	2 & $0.083333$ & $\frac{1}{12}$ & $6\times 3^{\frac{\L}{2}-1}$ \tabularnewline
	  & $0.750000$ & $\frac{3}{4}$ & $3\times 3^{\frac{\L}{2}-1}$ \tabularnewline 
	  & $0.750000$ & $\frac{3}{4}$ & $3\times 3^{\frac{\L}{2}-1}$ \tabularnewline \hline
\end{tabularx}
\end{table}

\noindent
We should remark that we have listed the conformal weight $h=\frac{3}{4}$
twice in Table \ref{tab:OpenExcitLow} to emphasize that there are two
different Bethe root configurations with different finite size energies
extrapolating to this conformal dimension for $\mathcal{L}\mod4 = 2$.

\subsubsection{Pairing rules and discussion}
This model is similar to the braided version and has the full global symmetry
of the algebra $D(D_{3})$. Every pair of solutions to the Bethe equations pair
which again implies that conformal weights will appear in multiple symmetry
sectors and the number of times they pair depends solely the symmetry sector
the eigenvalue of the transfer matrix lies in, Table
\ref{tab:PairingMultiplicity}.

\section{Discussion}
In this paper we have analysed the low energy spectrum of the integrable
$D(D_3)$ symmetric chain subject to various boundary conditions.  In the spin
chain formulation the Hamiltonian derives from a commuting two-parameter
transfer matrix of a vertex model and the eigenvalues can be obtained by Bethe
ansatz methods.
We have constructed a related class of models with local $D(D_3)$ symmetry
using the fusion path formulation: these models, too, are integrable as they
can be obtained from a solution to the Yang-Baxter equation for a face (or
RSOS) model.
For open and braided boundary conditions the two formulations of the model are
equivalent.  For periodic closure, however, the fusion path chain differs from
the spin chain by boundary terms.  Based on studies of small systems we have
proposed a set of Bethe equations whose solutions determine the eigenvalues of
the fusion path chain.

From a finite size analysis of the spectrum of these models we have identified
the conformal field theories providing an effective description of the low
energy modes to contain two sectors from -- depending on the parameter
$\theta$ -- the minimal model $\mathcal{M}_{(5,6)}$ (the three-state Potts
model) and the $Z_4$ parafermion.  The physical fields are products of
operators from these sectors.  The individual factors can carry fractional
(non-integer or non-(para)fermionic) spins implying the appearance of Virasoro
characters in the partition function of the model which have not been
discussed in the context of $\mathcal{M}_{(5,6)}$ or $Z_4$ alone.

The locality of physical fields in the model is guaranteed by pairing rules.
This situation is similar to other models with several gapless modes
propagating with different Fermi velocities
\cite{BoIK86,FraKor1990,FraYu1990}.  We have to emphasise, however, that in
the present models this factorisation of different modes is \emph{exact}
already for finite chains and on all energy scales, unlike in say the
separation of spin and charge degrees of freedom observed within the low
energy spectrum of the one-dimensional Hubbard model where the coupling
between the sectors becomes manifest in subleading corrections to scaling and
at higher energies.  Another difference for the model studied here is that the
two sectors of the effective theory are not related to subalgebras of the
global symmetry of the model.  Therefore, to establish the pairing rules and
corresponding multiplicities we have resorted to numerical studies of small
systems together with counting arguments for the total number of states of the
system.  For the spin chain formulation we found that the pairing is
determined by the boundary conditions and can be related to the residual
symmetry of a given eigenstate.  It is also reflected by the appearance of
infinite roots appearing in the configurations solving the Bethe equations of
the model.
The spectrum of the periodic $D(D_3)$ model in the fusion path formulation
also shows pairing on all energy scales.  Unlike in the spin chain
formulation, however, the pairing is not transitive and cannot be described
based on residual symmetries as in the spin chain.  The identification of the
pairing rules and as to whether these rules are connected to topological
invariants involving the $D(D_3)$ $F$-moves remains an interesting open
problem for this model.  

As a first step to address this problem, numerical methods could be used to
form conjectures.  Ultimately, however, it would be desirable to obtain the
complete picture starting from the integrable structures underlying this
model.  For the topological invariants this requires to relate them to
elements of the RSOS Yang Baxter algebra.  For the solution of the spectral
problem of the anyon chain the functional relations (\ref{eqnFunRelFusPath})
have to be established using the fusion procedure for RSOS transfer matrices.
We shall address these questions in future work.\\

\subsubsection*{{Acknowledgements}}
The authors would like to thank Jesper Romers for providing mathematica code
that generated the $F$-moves for the $D(D_{3})$ anyons, along with Karen
Dancer, Fabian Essler, Jon Links and Robert Weston for discussing various
topics relating to this article and referring the authors to relevant
literature. 

Parts of the numerical data used in this work have been obtained using the
RRZN cluster system at Leibniz Universit\"at Hannover.  Support for this
project by the Deutsche Forschungsgemeinschaft is gratefully acknowledged.\\

%\bibliographystyle{PeterStandard.bst}
%\bibliography{PeterReferences,base.bib,bound.bib}

\begin{thebibliography}{10}

\bibitem{Affl86}
 I. Affleck, \textit{Universal term in the free energy at a critical point 
 and and the conformal anomaly}, Phys. Rev. Lett. \textbf{56}, 746-748, 
 (1986).

\bibitem{AlDM92a}
G. Albertini, S. Dasmahapatra and B.M. McCoy, \textit{{S}pectrum and
  completeness of the integrable 3-state {P}otts model: a finite size study},
  Int. J. Mod. Phys. A, \textbf{7, Suppl. 1A}, 1-53, (1992).

\bibitem{AlDM92}
G. Albertini, S. Dasmahapatra and B.M. McCoy, \textit{{S}pectrum doubling and
  the extended {B}rillouin zone in the excitations of the three state {P}otts
  spin chain}, Phys. Lett. A, \textbf{170}, 397-403, (1992).

\bibitem{AlBB88}
F.C. Alcaraz, M.N. Barber and M.T. Batchelor, \textit{{C}onformal invariance,
  the {XXZ} chain and the operator content of two-dimensional critical
  systems}, Ann. Phys. (NY), \textbf{182}, 280-343, (1988).

\bibitem{AGLTT2011}
E. Ardonne, J. Gukelberger, A.W.W. Ludwig, S. Trebst and M. Troyer,
  \textit{Microscopic models of interacting Yang-Lee anyons}, New J. Phys.,
  \textbf{13}, 045006, (2011).

\bibitem{BaFiGi2002}
L. Balents, M.P.A. Fisher and S.M. Girvin, \textit{Fractionalization in an
  easy-axis Kagome antiferromagnet}, Phys. Rev. B, \textbf{65}, 224412, (2002).

\bibitem{BazRes1989}
  V.V. Bazhanov and N.Yu. Reshetikhin, \textit{Critical RSOS models and 
  conformal field theory}, Int. J. Mod. Phys. A, \textbf{04}, 115, (1989).
  
\bibitem{BoIK86}
N.M. Bogoliubov, A.G. Izergin and V.E. Korepin, \textit{{C}ritical exponents
  for integrable models}, Nucl. Phys. B, \textbf{275 [FS17]}, 687-705, (1986).
 
\bibitem{BlCN86}
  H.W.J. Bl{\"o}te, J.L. Cardy and M.P. Nightingale, \textit{Conformal 
  invariance, the central charge and universal finite-size amplitudes at 
  criticality}, Phys. Rev. Lett., \textbf{56}, 742-745, (1986).
 
\bibitem{CaIZ87}
A. Cappelli, C. Itzykson and J.B. Zuber, \textit{{M}odular invariant partition
  functions in two dimensions}, Nucl. Phys. B, \textbf{280 [FS18]}, 445-465,
  (1987).

\bibitem{Card86b}
J.L. Cardy, \textit{{E}ffect of boundary conditions on the operator content of
  two-dimensional conformally invariant theories}, Nucl. Phys. B, \textbf{275},
  200-218, (1986).

\bibitem{ChaPreBook1994}
V. Chari and A. Pressley, \textit{A guide to quantum groups}, Cambridge
  University Press, (1994).

\bibitem{DFIL2009}
K.A. Dancer, P.E. Finch, P.S. Isaac and J. Links, \textit{Integrable boundary
  conditions for a non-Abelian anyon chain with $D(D_{3})$ symmetry}, Nucl.
  Phys. B, \textbf{812}, 456--469, (2009).

\bibitem{DIL2006}
K.A. Dancer, P.S. Isaac and J. Links, \textit{Representations of the quantum
  double of finite group algebras and spectral parameter dependent solutions of
  the Yang--Baxter equation}, J. Math. Phys., \textbf{47}, 103511, (2006).

\bibitem{WilBai1998}
M. {de Wild Propitius} and F.A. Bais, \textit{Discrete gauge theories}, In
  Particles and Fields, Eds. G. Semenoff and L. Vinet, CRM Series in
  Mathematical Physics (Springer-Verlag, New York), 353--353, (1998).

\bibitem{FraZub1990}
P. Di~Francesco and J.B. Zuber, \textit{{${\rm SU}(N)$} lattice integrable
  models associated with graphs}, Nuclear Phys. B, \textbf{338}, 602--646,
  (1990).

\bibitem{DPR1990}
R. Dijkgraaf, V. Pasquier and P. Roche, \textit{Quasi Hopf algebras, group
  cohomology and orbifold models}, Nucl. Phys. (Proc. Supp.), \textbf{18},
  60--72, (1990).

\bibitem{EFGKVBook2005}
F. Essler, H. Frahm, F. G{\"o}hmann, A. Kl{\"u}mper and V.E. Korepin,
  \textit{The one-dimensional Hubbard model}, Cambridge University Press,
  (2005).

\bibitem{FatZam1982b}
V.A. Fateev and A.B. Zamolodchikov, \textit{Self-dual solutions of the
  star-triangle relation in $\mathbb{Z}_{N}$-Models}, Phys. Lett. A,
  \textbf{92}, 37--39, (1982).

\bibitem{FTLTKWF2007}
A. Feiguin, S. Trebst, A.W.W. Ludwig, M. Troyer, A.Y. Kitaev, Z. Wang and M.H.
  Freedman, \textit{Interacting Anyons in Topological Quantum Liquids: The
  Golden Chain}, Physical Review Letters, \textbf{98}, 160409, (2007).

\bibitem{Finch2011}
P.E. Finch, \textit{Integrable Hamiltonians with $D(D_n)$ symmetry from the
  Fateev-Zamolodchikov model}, J. Stat. Mech., P04012, (2011).

\bibitem{Finch2013}
P.E. Finch, \textit{From spin to anyon notation: The XXZ Heisenberg model as a
  $D_{3}$ (or $su(2)_{4}$) anyon chain}, J. Phys. A, 46, 055305, (2013).

\bibitem{FinFra2012}
P.E. Finch and H. Frahm, \textit{Collective states of interacting $D(D_3)$
  non-Abelian anyons}, J. Stat. Mech., \textbf{5}, L05001, (2012).

\bibitem{FFL2011}
P.E. Finch, H. Frahm and J. Links, \textit{Ground-state phase diagram for a
  system of interacting, $D(D_3)$ non-Abelian anyons}, Nucl. Phys. B,
  \textbf{844}, 129--145, (2011).

\bibitem{Foerster1996}
A. Foerster, \textit{Quantum group invariant supersymmetric {$t$}-{$J$} model
  with periodic boundary conditions}, J. Phys. A, \textbf{29}, 7625--7633,
  (1996).

\bibitem{FraKor1990}
H. Frahm and V.E. Korepin, \textit{Critical exponents for the one-dimensional
  Hubbard model}, Phys. Rev. B, \textbf{42}, 10553-10565, (1990).

\bibitem{FraYu1990}
H. Frahm and N.C. Yu, \textit{Finite size effects in the integrable XXZ
  Heisenberg model with arbitrary spin}, J. Phys. A, \textbf{23}, 2115, (1990).

\bibitem{Gepner1992}
D. Gepner, \textit{Foundations of Rational Quantum Field Theory, I}, Caltech
  preprint CALT-68-1825, arXiv:hep-th/9211100, (1992).

\bibitem{GeQi87}
D. Gepner and Z. Qiu, \textit{{M}odular invariant partition functions for
  parafermionic field theories}, Nucl. Phys. B, \textbf{285}, 423-453, (1987).

\bibitem{GATHLTW2013}
C. Gils, E. Ardonne, S. Trebst, D.A. Huse, A.W.W. Ludwig, M. Troyer and Z.
  Wang, \textit{Anyonic quantum spin chains: Spin-1 generalizations and
  topological stability}, arXiv:1303.4290, (2013).

\bibitem{GATLTW2009}
C. Gils, E. Ardonne, S. Trebst, A. Ludwig, M. Troyer and Z. Wang,
  \textit{Collective States of Interacting Anyons, Edge States, and the
  Nucleation of Topological Liquids}, Phys. Rev. Lett., \textbf{103}, 070401,
  (2009).

\bibitem{GRS1996}
C. G{\'o}mez, M. Ruiz-Altaba and G. Sierra, \textit{Quantum groups in
  two-dimensional physics}, Cambridge University Press, (1996).

\bibitem{Gould1993}
M.D. Gould, \textit{Quantum double finite group algebras and their
  representations}, Bull. Aust. Math. Soc., \textbf{48}, 275--301, (1993).

\bibitem{GPPR1994}
H. Grosse, S. Pallua, P. Prester and E. Raschhofer, \textit{On a quantum group
  invariant spin chain with non-local boundary conditions}, J. Phys. A: Math.
  Gen., \textbf{27}, 4761-4771, (1994).

\bibitem{Hamer81}
  C.J. Hamer, \textit{Q-state Potts models in Hamiltonian field theory for 
  $Qgeq 4$ in $(1+1)$ dimensions},J. Phys. A \textbf{14}, 2981-3003, (1981)
  
\bibitem{IkJS09}
Y. Ikhlef, J.L. Jacobsen and H. Saleur, \textit{{A} {T}emperley-{L}ieb quantum
  chain with two- and three-site interactions}, J. Phys. A, \textbf{42},
  292002, (2009).

\bibitem{KakArd2012}
P. Kakashvili and E. Ardonne, \textit{Integrability in anyonic quantum spin
  chains via a composite height model}, Phys. Rev. B, \textbf{85}, 115116,
  (2012).

\bibitem{KarZap1994}
M. Karowski and A. Zapletal, \textit{Quantum Group Invariant Integrable n-State
  Vertex Models with Periodic Boundary Conditions}, Nucl. Phys. B,
  \textbf{419}, 567-588, (1994).

\bibitem{Kitaev2006}
A. Kitaev, \textit{Anyons in an exactly solved model and beyond}, Ann. Phys.,
  \textbf{321}, 2--111, (2006).

\bibitem{Kita03}
A.{\relax Yu}. Kitaev, \textit{{F}ault-tolerant quantum computation by anyons},
  Ann. Phys. (NY), \textbf{303}, 2-30, (2003).

\bibitem{KKMST2011}
N. Kitanine, K.K. Kozlowski, J.M. Maillet, N.A. Slavnov and V. Terras,
  \textit{A form factor approach to the asymptotic behavior of correlation
  functions in critical models}, J. Stat. Mech., P12010, (2011).

\bibitem{KBIBook1993}
V.E. Korepin, N.M. Bogoliubov and A.G. Izergin, \textit{Quantum inverse
  scattering method and correlation functions}, Cambridge University Press,
  (1993).
  
\bibitem{KluPea1992}
A. Kl{\"u}mper, N.M. and P.A Pearce, \textit{Conformal weights of RSOS lattice
  models and their fusion hierarchies}, Physica A: Statistical Mechanics and 
  its Applications, \textbf{183}, 304-350, (1992).
  
  
\bibitem{Laughlin1983}
R.B. Laughlin, \textit{Anomalous quantum Hall effect: an incompressible quantum
  fluid with fractionally charged excitations}, Phys. Rev. Lett., \textbf{50},
  1395, (1983).

\bibitem{LPTT2011}
A.W. Ludwig, D. Poilblanc, S. Trebst and M. Troyer, \textit{Two-dimensional
  quantum liquids from interacting non-Abelian anyons}, New J. Phys.,
  \textbf{13}, 045014, (2011).

\bibitem{MajidBook1995}
S. Majid, \textit{Foundations of quantum group theory}, Cambrigde University
  Press, (1995).

\bibitem{CoyKed1993}
B.M. McCoy and R. Kedem, \textit{Construction of modular branching functions
  from Bethe's equations in the 3-state Potts chain}, J. Stat. Phys,
  \textbf{17}, 865, (1993).

\bibitem{McKay1980}
J. McKay, \textit{Graphs, singularities, and finite groups}, The {S}anta {C}ruz
  {C}onference on {F}inite {G}roups ({U}niv. {C}alifornia, {S}anta {C}ruz,
  {C}alif., 1979), \textbf{37}, 183--186, (1980).

\bibitem{MoeSon2001}
R. Moessner and S.L. Sondhi, \textit{Resonating valence bond phase in the
  triangular lattice quantum dimer model}, Phys. Rev. Lett., \textbf{86}, 1881,
  (2001).

\bibitem{MooRea1991}
G. Moore and N. Read, \textit{Nonabelions in the fractional quantum Hall
  effect}, Nucl. Phys. B, \textbf{360}, 362--396, (1991).

\bibitem{NSSF08}
C. Nayak, S.H. Simon, A. Stern, M. Freedman and S.D. Sarma,
  \textit{{N}on-{A}belian {A}nyons and {T}opological {Q}uantum {C}omputation},
  Rev. Mod. Phys., \textbf{80}, 1083-1159, (2008).

\bibitem{Pasquier1987b}
V. Pasquier, \textit{Lattice derivation of modular invariant partition
  functions on the torus}, J. Phys. A, \textbf{20}, L1229--L1237, (1987).

\bibitem{Pasquier1987a}
V. Pasquier, \textit{Two-Dimensional Critical Systems Labelled by Dynkin
  Diagrams}, Nucl. Phys. B, \textbf{285}, 162--172, (1987).

\bibitem{Pasquier1988}
V. Pasquier, \textit{Etiology of {IRF} models}, Comm. Math. Phys.,
  \textbf{118}, 355--364, (1988).

\bibitem{Roche1992}
P. Roche, \textit{On the construction of integrable dilute ADE models}, Physics
  Letters B, \textbf{285}, 49-53, (1992).
  
\bibitem{RomersPrivate}
J. Romers, \textit{Private communication} (2011).

\bibitem{Sklyanin1988}
E.K. Sklyanin, \textit{Boundary conditions for integrable quantum systems}, J.
  Phys. A: Math. Gen., \textbf{21}, 2375-2389, (1988).

\bibitem{Suzu88}
J. Suzuki, \textit{{S}imple excitations in the nested {Bethe}-ansatz model}, J.
  Phys. A: Math. Gen., \textbf{21}, L1175--L1180, (1988).

\bibitem{TAFHLT2008}
S. Trebst, E. Ardonne, A. Feiguin, D. Huse, A. Ludwig and M. Troyer,
  \textit{Collective states of interacting Fibonacci anyons}, Phys. Rev. Lett.,
  \textbf{101}, 050401, (2008).

\bibitem{TTWL2008}
S. Trebst, M. Troyer, Z. Wang and A.W.W. Ludwig, \textit{A short introduction
  to Fibonacci anyon models}, Prog. Theor. Phys. Suppl., \textbf{176}, 384,
  (2008).

\bibitem{GehRit1986}
G. {von Gehlen} and V. Rittenberg, \textit{Operator content of the three-state
  Potts quantum chain,}, J. Phys. A: Math. Gen., \textbf{19}, L625, (1986).

\bibitem{WarNie1993}
S.O. Warnaar and B. Nienhuis, \textit{Solvable lattice models labelled by
  {D}ynkin diagrams}, J. Phys. A, \textbf{26}, 2301--2316, (1993).

\bibitem{YaYa69}
C.N. Yang and C.P. Yang, \textit{{T}hermodynamics of a one-dimensional system
  of bosons with repulsive delta-function interaction}, J. Math. Phys.,
  \textbf{10}, 1115-1122, (1969).

\bibitem{ZaFa85}
A.B. Zamolodchikov and V.A. Fateev, \textit{{N}onlocal (parafermion) currents
  in two-dimensional conformal quantum field theory and self-dual critical
  points in {$Z_n$}-symmetric statistical systems}, Sov. Phys. JETP,
  \textbf{62}, 215-225, (1985).

\end{thebibliography}

\begin{appendix}
\section{Dressed charge formalism}
Following \cite{BoIK86,Suzu88,FraYu1990,EFGKVBook2005} the finite size energy
gaps of $\mathcal{H}_{\theta=0}$ for periodic boundary conditions are
\begin{equation}
\begin{aligned}
  \Delta E(\Delta N_\pm,Q_\pm) = \frac{2\pi}{\mathcal{L}} v_F &\Big(
\frac{1}{4} (\Delta N_+,\Delta N_-) (\Xi^\top \Xi)\,^{-1}
(\Delta N_+,\Delta N_-)\\
&  + (Q_+,Q_-) (\Xi^\top \Xi) (Q_+,Q_-) + \mathcal{N}
\Big) + o(\mathcal{L}^{-1})\,.
\end{aligned}
\end{equation}
($\mathcal{N}$ being a non-negative integer).  Taking into account that only
$\pm$-strings are present in the ground state the $2\times2$ dressed charge
matrix $\Xi =\xi(x)|_{x=\infty}$ is obtained from the linear integral equation
\begin{equation}
\begin{aligned}
  \xi(x) &= \left(\begin{array}{cc}1&0\\0&1
    \end{array}\right) - \frac{1}{2\pi}\int_{-\infty}^{\infty} \mathrm{d}y\,
  \xi(y) K(y-x)\,,\\
  K(x) &= \left(\begin{array}{cc}
      k(x,\frac{1}{3})&k(x,\frac{5}{6})\\k(x,\frac{5}{6})&k(x,\frac{1}{3})
    \end{array}\right)\,,\quad
  k(x,t) = \frac{\sin(2\pi t)}{\cosh(x)-cos(2\pi t)}.
  \end{aligned}
\end{equation}
Using Wiener Hopf techniques the dressed charge matrix can be expressed in
terms of the Fourier transform of the kernel matrix giving
\begin{equation}
  \Xi^\top \Xi = \left(1-\widetilde{K}(\omega=0)\right)^{-1} = 
  \left( \begin{array}{cc} 1 & \frac{1}{2}\\\frac{1}{2} & 1
    \end{array}\right)\,.
\end{equation}
Hence the scaling dimensions and conformal spins of primary operators in the
effective field theory for $\mathcal{H}_{\theta=0}$ in terms of the quantum
numbers $\Delta N_\pm$ and $Q_\pm$ characterising the corresponding excitation
(\ref{cft}) are
\begin{equation}
\label{scgap2}
\begin{aligned}
  X & = \frac{1}{3}\left((\Delta N_{+})^{2} - \Delta N_{+} \Delta N_{-}
    + (\Delta N_{-})^{2} \right) + \left( (Q_{+})^{2} + Q_{+}  Q_{-} +
    (Q_{-})^{2} \right)\,, \\ 
  s & = -\frac{1}{2}\left(Q_{+}\Delta N_{+} + Q_{-}\Delta
    N_{-}\right)\,.
\end{aligned}
\end{equation}
The $\Delta N_{\pm}\equiv N_{\pm} - \frac{\mathcal{L}}{2} \pm \frac{\mathcal{L}}{4}$ correspond
to the change in number of $\pm$-strings (subject to the condition that the
total number of roots is $\mathcal{L}$) as compared to the thermodynamic
ground state while the $Q_{\pm}$ determine the momentum of the excitation.
For a configuration with $n_\pm$ Bethe roots at $\pm\infty$ they can take
discrete values
$$ Q_{\pm} \equiv  \frac{1}{3}(n_{+\infty}-n_{-\infty}) \quad (\mbox{mod 1}). $$
For a given solution $\{x_k^\pm\}$ of the Bethe equations the $Q_\pm$ can also
be determined numerically using the counting functions defined above:
$$  Q_{\pm} = \frac{1}{N_{\pm}} \sum_{k=1}^{N_{\pm}} Z_{\pm}(x_{k}^{\pm})\,. $$ 
Due to the discrete set of possible values for $Q_\pm$ the data for systems
with $\mathcal{L}\le10$ are sufficient to identify the quantum numbers for the
lowest finite size gaps of $\mathcal{H}_{\theta=0}$, see in Table
\ref{tab:ExPerLowAltFull}.  The observed dimensions support our identification
of the critical theory with a $Z_4$ parafermion.

\begin{table}[ht]
  \caption{\label{tab:ExPerLowAltFull}The lowest excitations of $\mathcal{H}_{0}$ in terms of the quantities $\Delta N_{\pm}$ and $Q_{\pm}$ for different chain lengths.}
\begin{tabularx}{\textwidth}{CCCCCCC} \hline \hline
	$\mathcal{L}$ mod 4 & $X$ & $s$ & $\Delta N_{+}$ & $\Delta N_{-}$ & $\Delta Q_{+}$ & $\Delta Q_{-}$ \tabularnewline \hline \hline
	0 & $0$ & $0$ & 0 & 0 & 0 & 0 \tabularnewline
	  & $\frac{1}{3}$ & $-\frac{1}{3}$  & 0 & -1 & $-\frac{1}{3}$ & $+\frac{2}{3}$ \tabularnewline 
	  & $\frac{2}{3}$ & $0$  & 0 & -2 & 0 & 0 \tabularnewline
	  & $\frac{1}{6}$ & $0$  & -1 & -1 & 0 & 0 \tabularnewline \hline
	1 & $\frac{1}{16}$ & $\frac{1}{16}$ & $-\frac{1}{4}$ & $+\frac{1}{4}$ & $+\frac{1}{4}$ & $-\frac{1}{4}$ \tabularnewline
	  & $\frac{7}{48}$ & $-\frac{1}{48}$ & $-\frac{1}{4}$ & $-\frac{3}{4}$ & $-\frac{5}{12}$ & $+\frac{1}{12}$ \tabularnewline
	  & $\frac{19}{48}$ & $-\frac{13}{48}$ & $-\frac{1}{4}$ & $-\frac{3}{4}$ & $+\frac{7}{12}$ & $-\frac{11}{12}$ \tabularnewline \hline
	2 & $\frac{1}{12}$ & $-\frac{1}{12}$  & $-\frac{1}{2}$ & $-\frac{1}{2}$ & $-\frac{1}{6}$ & $-\frac{1}{6}$ \tabularnewline
	  & $\frac{5}{12}$ & $-\frac{1}{4}$  & $-\frac{1}{2}$ & $-\frac{3}{2}$ & $-\frac{1}{2}$ & $+\frac{1}{2}$ \tabularnewline \hline \hline
\end{tabularx}
\end{table}

\end{appendix}

\end{document}